\documentclass[10pt]{extarticle}
\usepackage[a4paper, margin=1.5cm]{geometry}
\usepackage{authblk}
\usepackage{graphicx}
\usepackage{listings}
\usepackage{xcolor}%
\usepackage{amsmath}
\usepackage{amssymb}
\usepackage{appendix}
\usepackage{float} 
\DeclareMathOperator{\arctanh}{arctanh}
\usepackage{subcaption}
\usepackage{minted}
\usepackage{graphicx}
\title{Inferred and Residual Parameter Interactions in Syntax Dynamics}
\author[1]{Hediye Yarahmadi\thanks{hediye.yarahmadi@gmail.com}}
\author[1]{Alessandro Treves\thanks{ale@sissa.it}}
\affil[1]{SISSA - Cognitive Neuroscience, Trieste, Italy}
\date{\today}
\usepackage[colorlinks=true, linkcolor=blue, citecolor=blue, urlcolor=blue]{hyperref}

\begin{document}

\maketitle

\begin{abstract}
    What makes natural languages diversify, in their syntactic structure, even when sharing a largely common origin? We refine our simple model, which focuses on the glassy dynamics induced by a network of interactions between syntactic parameters. Distinguishing between symmetric interaction weights that can be inferred from the observed statistics of syntax across many languages, and additional asymmetric weights assigned at random, we show that both components are needed, with similar overall strength, to account for plausible long-term syntactic dynamics. An approximate balance can combine the glassy metastability provided by the inferred weights with the chaotic diversification induced by the residual asymmetric weights.
\end{abstract}

\section{Introduction and objective}

Why have the natural languages of the world, which change in the course of time, not evolved towards a common, optimally efficient form? The question is particularly relevant to the core internal structure of languages, their syntax. In \cite{yarahmadi2026modeling} we have taken advantage of the Principles and Parameters description of syntax in theoretical linguistics \cite{chomsky1993theory} and modeled the evolution of syntactic parameters as determined, within a closed system isolated from outside influences, by randomly assigned interaction weights. We have found that both in a purely binary model, in which parameters only take $\pm 1$ values, and in a model where parameters can also remain undefined -- they stay in a 0-state if the relevant `implication' is not satisfied -- evolution dynamics are glassy for symmetric interactions and up to a critical value $\phi_c =30^{\circ} $ of the asymmetry angle. For larger asymmetry, dynamics are fluid. Glassy dynamics are plagued by `frustration', the inability of the system to find states fully coherent with all its biases \cite{mezard1987spin}. This might explain what makes syntax linger around locally quasi-satisfactory states, rather than converge towards a global optimum. Are such metastable states, however, numerous enough to account for the observed diversity across natural languages?

\begin{figure}
\centering
        \includegraphics[width=15cm]{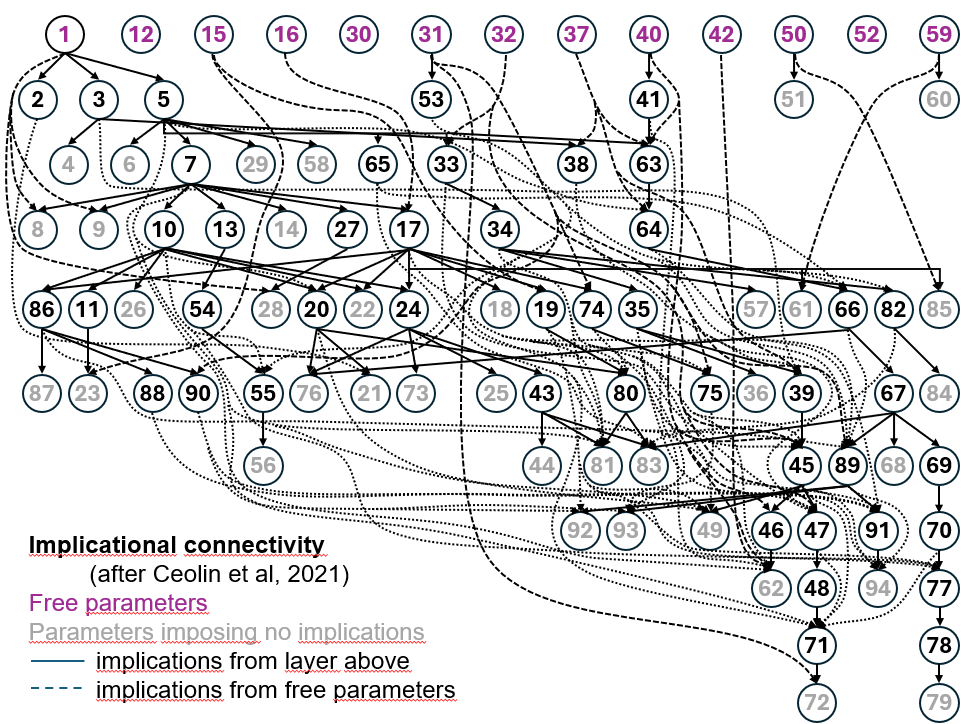}
    \caption{Syntactic implications operate at different depth levels, or layers, $l=0,\dots,10$, top to bottom, depending on where they are embedded in this `mangrove' diagram, shown as arrows pointing at the dependent parameters, which are all numbered as in the PCM database \cite{ceolin2021boundaries}. Boolean rules are not represented in this simple connectivity diagram (but are detailed in their Supplemental Table 2).}
    \label{fig:implications}
\end{figure}

The key idea now is that to proceed further and eventually model the evolution of the syntax of specific languages, interaction weights should not be completely random, but rather reflect the observed syntax statistics. After all, they should reflect the interactions, in the brain of the speakers, that contribute to those statistics. For example, if a +1 value for parameter A is logically or linguistically incompatible with a +1 value for parameter B, this will be expressed by an appropriate implication; but if it is just difficult for the brain to organize its syntactic schemata such that both parameters are set to +1, this can be expressed by a (negative) interaction weight, which should be reflected in observing fewer languages, with those values, than expected by chance. Weights in the network model should therefore include, alongside a random component, a statistics-inferred component of relative strength $\zeta$ ($\zeta $ thus quantifies, as in \cite{yarahmadi2026modeling}, the syntax-directed departure from a structure-less statistical model). Such inference, again, accrues on top of the observation that syntactic parameters are not linguistically independent of each other; rather, they are embedded in a hierarchy of so-called {\em implications}, which determine when a specific parameter can be defined or else is forced to stay in the 0-state. For the PCM database we use, which concerns only syntactic parameters pertaning to the nominal phrase \cite{ceolin2021boundaries}, the list of implications is represented graphically in Fig.\ref{fig:implications} \cite{yarahmadi2026modeling}. We aim here to determine whether, when adding the statistics-inferred component, glassy dynamics prevail in a similar region of the phase diagram, as in \cite{yarahmadi2026modeling}; and if so, how diverse are the trajectories followed by different languages. Moreover, we ask what the statistics of existing natural languages can tell us about their past and future, within this simple modeling framework. These objectives are pursued here only with the refined model that includes the 0-state, in which parameters are first determined as defined or undefined and only if defined are subject to zero-temperature updating dynamics, as described below.

\section{Modelling Methods}

We assume the syntactic parameters considered in 
\cite{ceolin2021boundaries} to comprise an interacting network, that is, each parameter evolves subject to interactions with the other parameters. This proceeds, in our simplified model, over `time steps'. At each time step, going down the implicational hierarchy in Fig.\ref{fig:implications}, first the implication-free parameters at the top of the hierarchy are set to either +1 or -1, then, level by level, it is determined which of the parameters at that level are defined, based on the values of those in the levels above, and if so are set to either +1 or -1. Setting a parameter value to $\pm 1$ is subject to a multitude of effects we cannot conceivably account for in detail. The idea is to focus on gross statistical imbalances between the two values, expressed by suitable {\em thresholds}, $\theta$, and on the interactions with the other parameters considered in \cite{ceolin2021boundaries}, expressed by a weight matrix $J$, and to ignore everything else. It can be taken as a peculiar type of {\em Signal} vs {\em Noise} distinction, where in the latter category we first dump a multitude of effects both `fast' and `slow', including e.g. interactions with many other syntax parameters (such as those of the {\em verb phrase} not considered in \cite{ceolin2021boundaries}), with other linguistic and socio-geographic variables, etc.; and then we pretend the noise variables to vanish. Thus, for both the fast noise, which could be modeled as an effective temperature $T\equiv 1/\beta $, and the slow noise, which could be expressed by an additional variable $\eta_i$, we consider only the limit case $T\to 0, \quad {\rm var}(\eta) \to 0$ in the update dynamics
\begin{equation}
    S_i = [{\rm if \; defined}] = {\rm sign} \left[ \sum_j J_{ij} S_j + \theta_j \right].
    \label{updating}
\end{equation}
This simple equation is not a statement about the mechanics of linguistic change, but rather an attempt to gain insight from what one can extract or infer from the available statistics, i.e. thresholds, affecting individual parameters, and pairwise interaction weights, determining how parameters affect each other.

\subsection{Thresholds}

As the network of parameters is assumed in this model to be isolated, the only terms determining the $\pm 1$ values of defined parameters are the weights and thresholds, and the overall amplitude of the former can be reabsorbed into the definition of the latter. 

Abstracting from the interactions with the other parameters, a parameter $i$, if defined, takes values $S_i =\pm 1$ in the model depending on its threshold $\theta_i$. We estimate the basic probability (or underlying frequency, $f_i$) with which this occurs from the number $n(S_i=\pm 1)$ of natural languages in the PCM database \cite{ceolin2021boundaries} in which it takes either $+1$ or $-1$ value, regularized
\begin{equation}
    f_i^+ = \frac{n(S_i=+1)+0.5}{n(S_i=+1)+n(S_i=-1)+1}
\end{equation}
where the +0.5, +1 terms enforce a moderate regularization (a weaker version of Laplace's Rule of Succession \cite{zabell1989rule}, as we have no basis to assume a uniform prior) and prevent infinities arising from cases of $f=0$ or 1, see App. A. We then set
\begin{equation}
\theta_i=\arctanh (1-2f_i^+). 
\label{theta_set}
\end{equation}
In this first formulation of the model, we have no random component in the threshold, so that the conditional probability is indeed, when considering the interactions with other parameters as a form of slow noise,
\begin{equation}
    {\rm Prob}(S_i=+1|S_i \;{\rm is}\;{\rm defined})= (1/2)[1-\tanh{\theta_i}]=f^+_i.
\end{equation}

\subsection{Couplings}

The underlying frequency with which two parameters $i$ and $j$ take, e.g., joint values $S_i=+1$ and $S_j=-1$, if both are defined, is similarly estimated as 
\begin{equation}
    f_{ij}^{+-}=\frac{n(S_i=+1,S_j=-1)+0.25}{n(+1,+1)+n(+1,-1)+n(-1,+1)+n(-1,-1)+1},
\end{equation}
and we assume that it deviates from the product of the single-parameter frequencies mainly due to a single interaction term
\begin{equation}
    f_{ij}^{+-}\simeq f_i^+ f_j^-\exp{(- w^I_{ij}/2)},
\end{equation}
where the superscript $I$ indicates that this component of the couplings can be {\em inferred} from the statistics of the parameter vectors in natural languages. We then note that $w^I_{ij}$ could be estimated in four different ways 
\begin{equation}
    \exp{( w^I_{ij}/2)} \quad 
    \simeq \frac{f_{ij}^{++}}{f_{i}^{+}f_{j}^{+}};
    \quad \simeq \frac{f_{ij}^{--}}{f_{i}^{-}f_{j}^{-}};
    \quad \simeq \frac{f_{i}^{+}f_{j}^{-}}{f_{ij}^{+-}};
    \quad \simeq \frac{f_{i}^{-}f_{j}^{+}}{f_{ij}^{-+}}
\end{equation}
and we take the simple average (i.e. the geometric average of the exponential) and define
\begin{equation}
    w^I_{ij} \equiv \frac{1}{2}\ln 
    \frac{f_{ij}^{++}f_{ij}^{--}}{f_{ij}^{+-}f_{ij}^{-+}}.
    \label{w-no-alpha}
\end{equation}

\begin{figure}
        \includegraphics[width=0.475\textwidth]{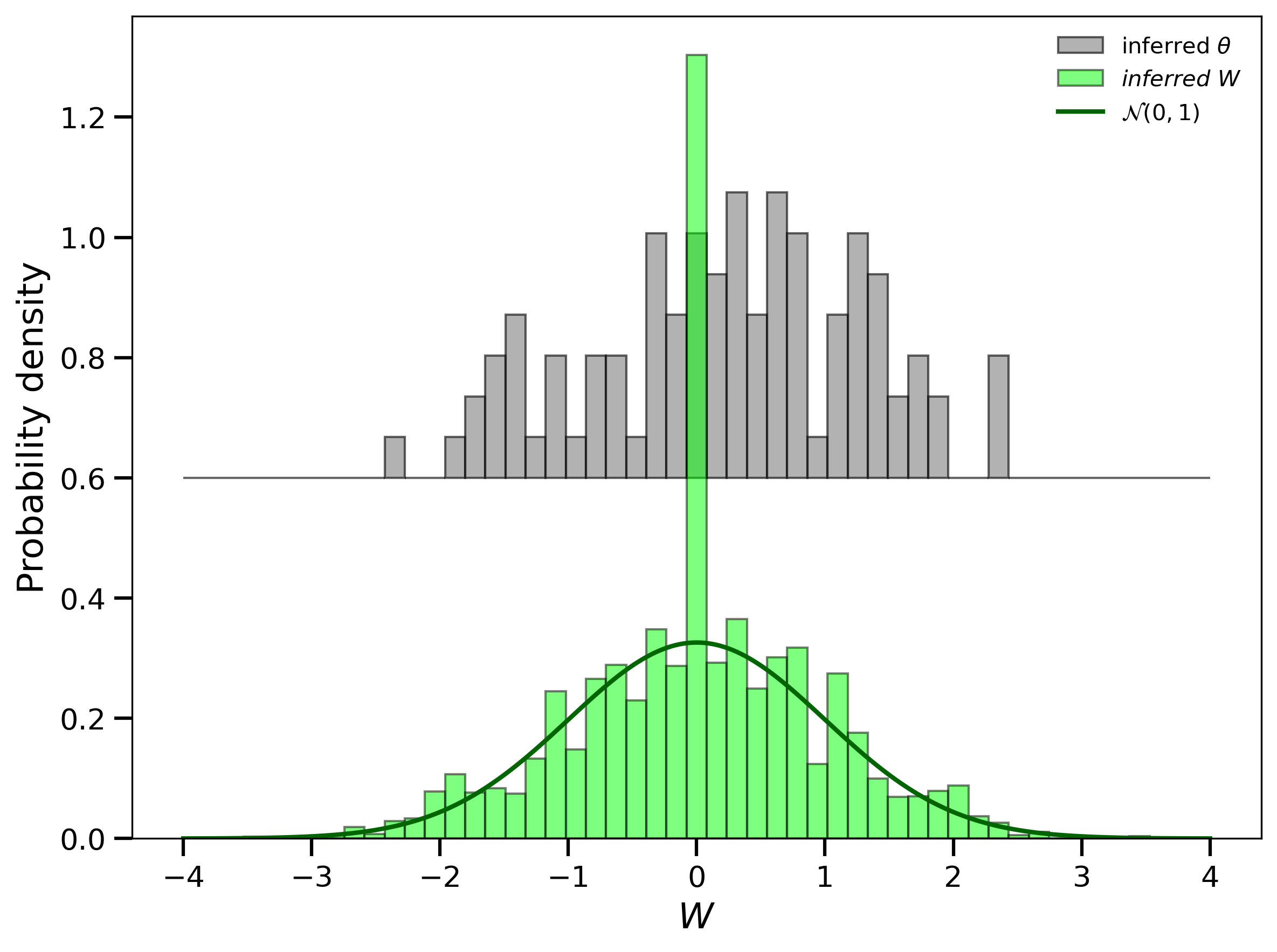}
\hspace{\fill}
        \includegraphics[width=0.475\textwidth]{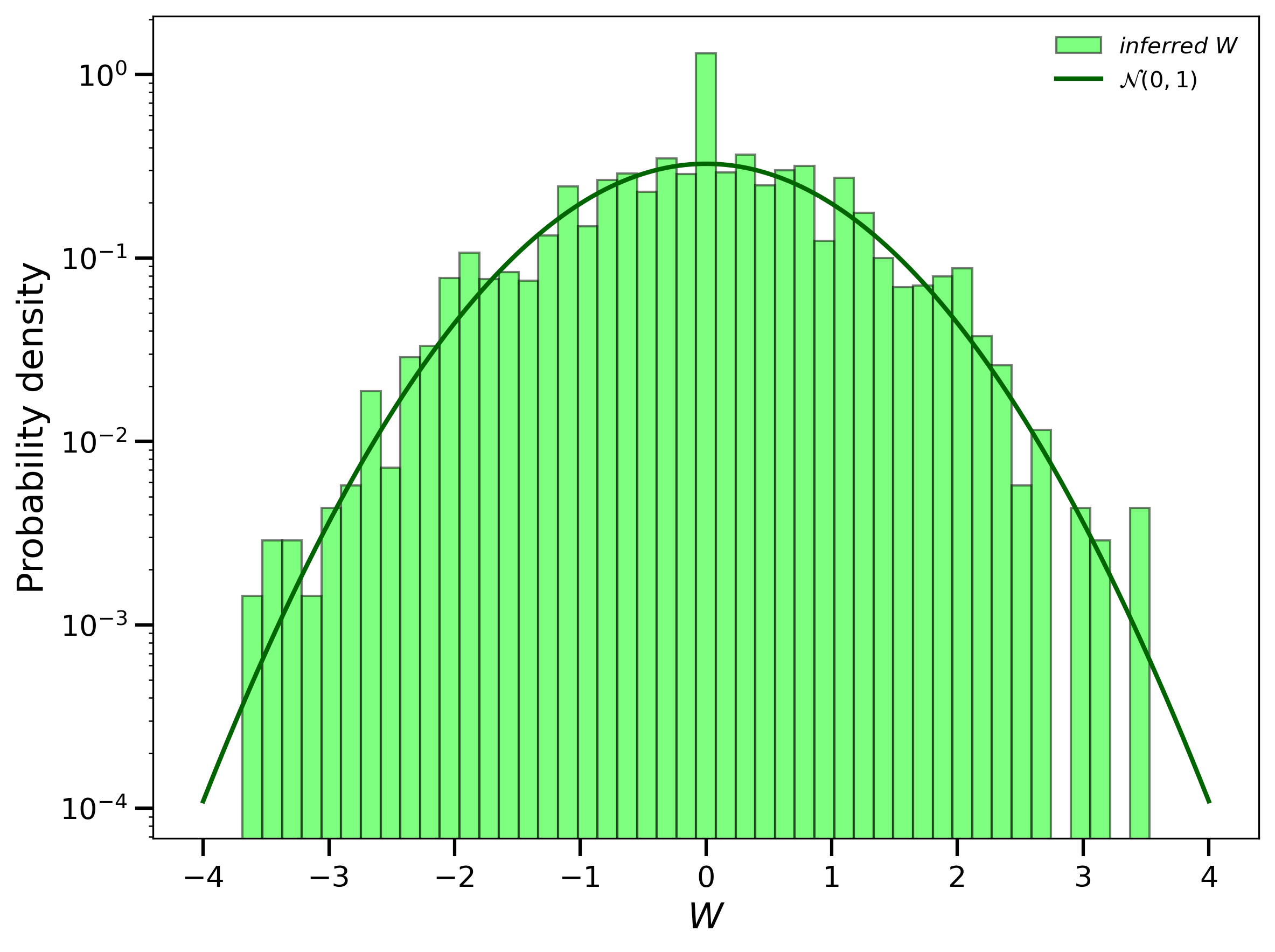}
    \caption{Distribution of $w^I_{ij}$ for different values, shown on normal/linear (left histogram) and logarithmic (right) scales, with a normal distribution (curves) of zero mean and unit variance, of amplitude equal to the fraction of non-zero inferred weights; the left panel also shows the distribution of the 94 inferred thresholds (too few to indicate in this case a model distribution). }
    \label{fig:alpha_comparison}
\end{figure}

This formula, Eq.(\ref{w-no-alpha}), is symmetric in $i,j$ and reflects the fact that if two parameters
tend to be in the same state across languages (both positive or both negative) more than predicted by chance (i.e., by their thresholds), this indicates a positive interaction between them. A purely random assignment of weights would fail to reproduce such statistical dependencies, so the model interactions must include structured ground/baseline values that capture these correlations.
We plot the distribution of $w^I_{ij}$ and $\theta_i$ values (Eqs.~\ref{w-no-alpha} and ~\ref{theta_set}) in Fig.\ref{fig:alpha_comparison}. 

The $w^I_{ij}$ distribution contains a number of zero values because, in some cases, the two parameters are not simultaneously defined in any of the languages being compared. In such cases, $w^I_{ij}$ is set to zero. Excluding the remaining zero-valued weights, the inferred weights closely follow a normal distribution, similar to the residual weights that will now be assigned in the model.

Besides the weights inferred from the statistics observed in the database, we add a {\em r}esidual contribution $W^R_{ij}$ to the overall couplings $J_{ij}$, to represent a multitude of other effects, including all the languages not present in the database, as well as the potential corrections induced e.g. by weighing the languages by the number of speakers, or by history-dependent factors
\begin{equation}
J_{ij}
=w^I_{ij} +\frac{1}{\zeta}W^R_{ij}=
\frac{1}{2}\ln 
    \frac{f_{ij}^{++}f_{ij}^{--}}{f_{ij}^{+-}f_{ij}^{-+}}
 +\frac{1}{\zeta}W^R_{ij}.
 \label{Jij}
\end{equation}
We consider only couplings between parameters which do interact in the database, i.e., both parameters are defined in at least one of those natural languages, and leave $J_{ij} =0$ if in none of those languages they are both defined.

The residual components are randomly assigned, further decomposed into symmetric and antisymmetric contributions as in \cite{yarahmadi2026modeling}
\begin{equation}
W^R_{ij}=\cos(\phi)W^S_{ij}
+
\sin(\phi)W^A_{ij},
\end{equation}
where
\[
W^S_{ij}=W^S_{ji},
\qquad
W^A_{ij}=-W^A_{ji},
\]
and both $W^S_{ij}$ and $W^A_{ij}$ are random variables drawn from a Gaussian 
distribution with zero mean and unit variance
\[
\mathrm{Var}(W^S_{ij})
=
\mathrm{Var}(W^A_{ij})
=1,
\]
so that the relative importance of inferred to residual components is regulated solely by $\zeta$.

The angle $\phi$ controls, instead, the relative contribution of antisymmetric to symmetric components in the residual weights. An angle $\tilde{\phi}$ can be similarly defined to quantify the degree of asymmetry of the full couplings in Eq. \ref{Jij}, as shown in App. \ref{app:asymm}.

\section{Results}
To characterize the dynamics of the parameter network, we run multiple simulations, each starting from one of the 58 `natural languages' (i.e., their syntactic parameter vectors) in the PCM database. Besides averaging over the starting languages, when appropriate we average over $N_r$ runs with the same random set of residual weights, and then over $N_w$ different sets. A simulation runs for up to $t=100$ `time steps' (but in general in the glassy regime parameters stop evolving much sooner) and in each time step we apply the simple updating in Eq.\ref{updating}, going down the levels of the implication `mangrove' in Fig.\ref{fig:implications}. Within each level, the updating order is random, and this is the only source of randomness in simulations with the same starting point and set of weights, as we do not consider other sources of fast noise.

\subsection{Overall glassiness}
An overall statistical analysis is based on 250 simulations per language, comprising $N_w=50$ distinct sets of residual weights, each evaluated over $N_r=5$ independent runs. We 
extract the average number of parameters that change their values, e.g., at time steps 1, 2 and 100, and 
plot these numbers as a function of the asymmetry angles $\phi$ and $\tilde{\phi}(\phi, \zeta)$ (see the figures below). 
The results are presented separately for parameters that enter or leave the 
$0$ state and for those that flip between the values $\pm 1$. Here, $\tilde{\phi}$ represents the degree of asymmetry of the full couplings, while $\phi$ is the degree of asymmetry of the residual couplings alone.

\begin{figure}
\centering
\includegraphics[width=0.49\linewidth]{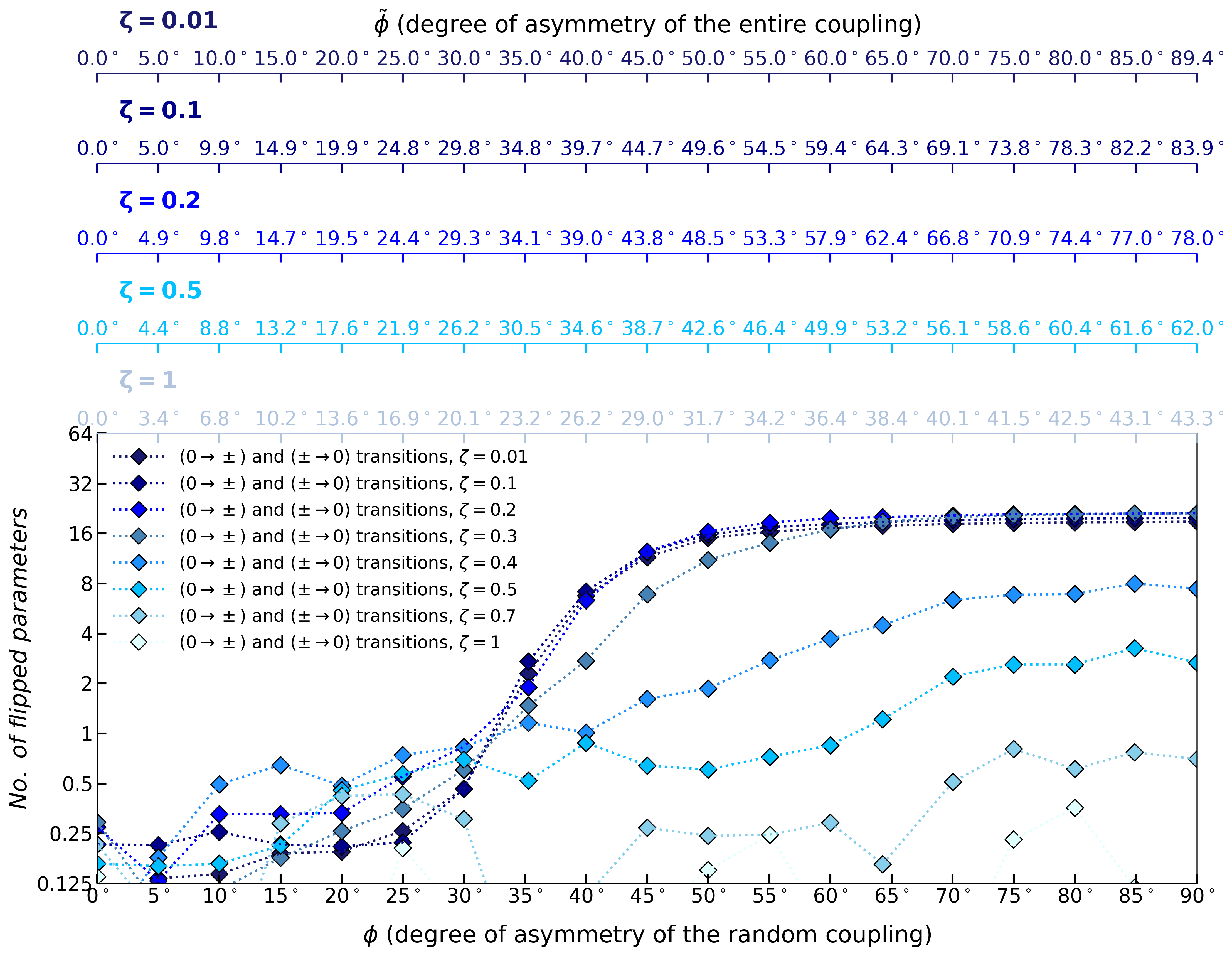}
\hspace{\fill}
\includegraphics[width=0.49\linewidth]{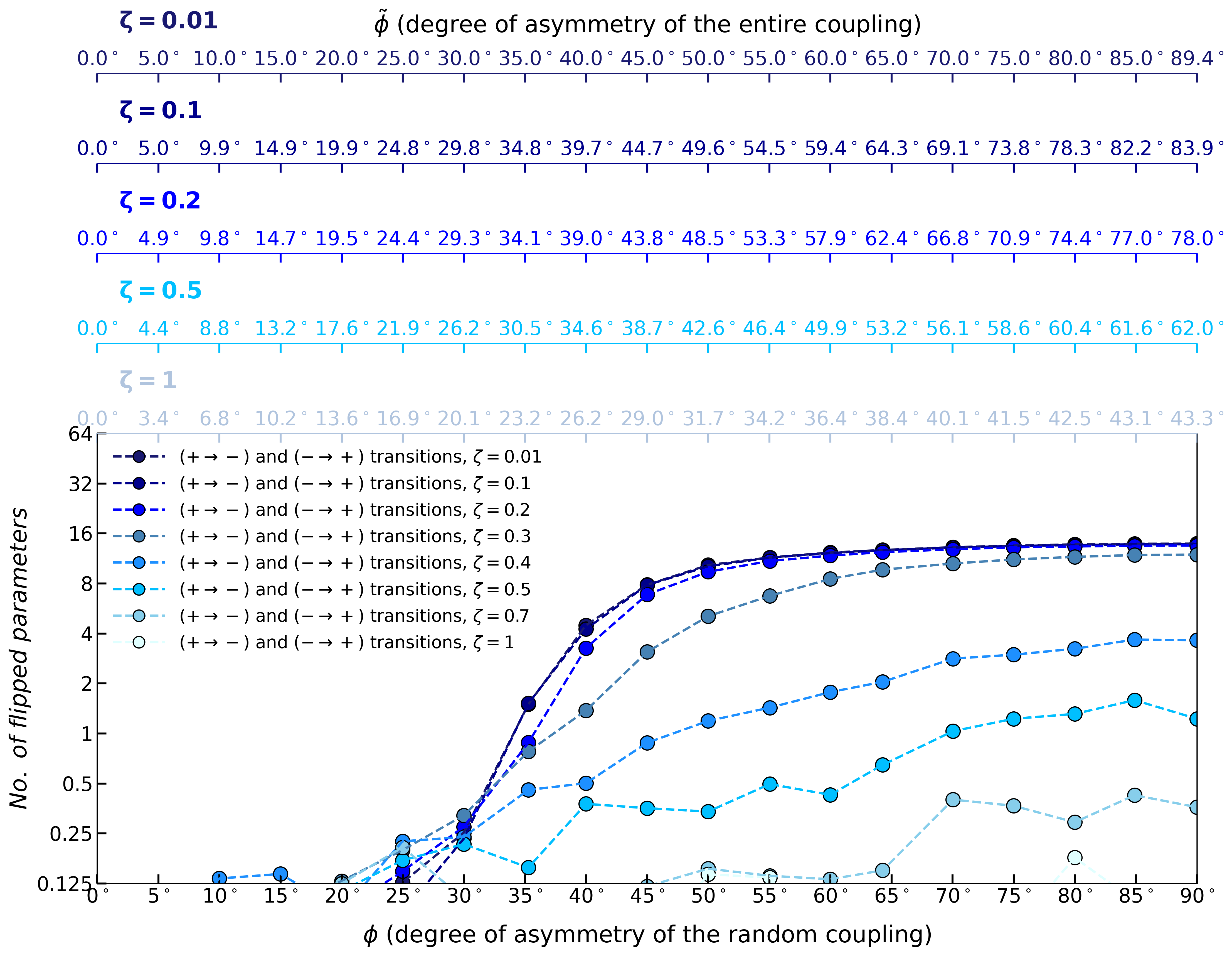}
\caption{The number of parameters that change from defined to undefined or viceversa (left) and of those that flip between their $\pm 1$ values (right), at time step 100, averaged across simulations, as a function of $\phi$ and $\zeta$ (and therefore of $\tilde{\phi}$). Note the logarithmic {\em y}-scale, with data points $<$ 0.125 not shown.}
\label{fig:fig2}
\end{figure}

Fig.\ref{fig:fig2} presents the results for different values of $\zeta$. The results indicate that $\zeta$ has a major influence on the system behavior. Only for small $\zeta$ one observes a robust transition from glassy to fluid dynamics near $\phi_c=30^{\circ}$, as in \cite{yarahmadi2026modeling}. When $\zeta$ is large, in fact $\zeta \gtrsim 1$ suffices, the effect of the residual weights becomes small and the dynamics remain glassy, irrespective of the degree of symmetry of those weights. Specifically, when $\zeta = 1$, for values $\phi\gtrsim 47^{\circ}$ one would have $\tilde{\phi}>30^{\circ}$, that is, the full couplings should be sufficiently asymmetric to make the parameter system fluid; however, the network still gets essentially stuck, with much less than one parameter flipping in any direction by $t=100$, on average.

In App. \ref{app:regul} we show that the exact regularization adopted does not affect the basic glassiness of the regime prevailing for $\zeta =1$. We then use the `standard' regularization detailed in the text for the rest of the analyses.

\begin{figure}[htbp]
    \centering
    \begin{subfigure}{8.5cm}
        \centering
        \includegraphics[width=\textwidth]{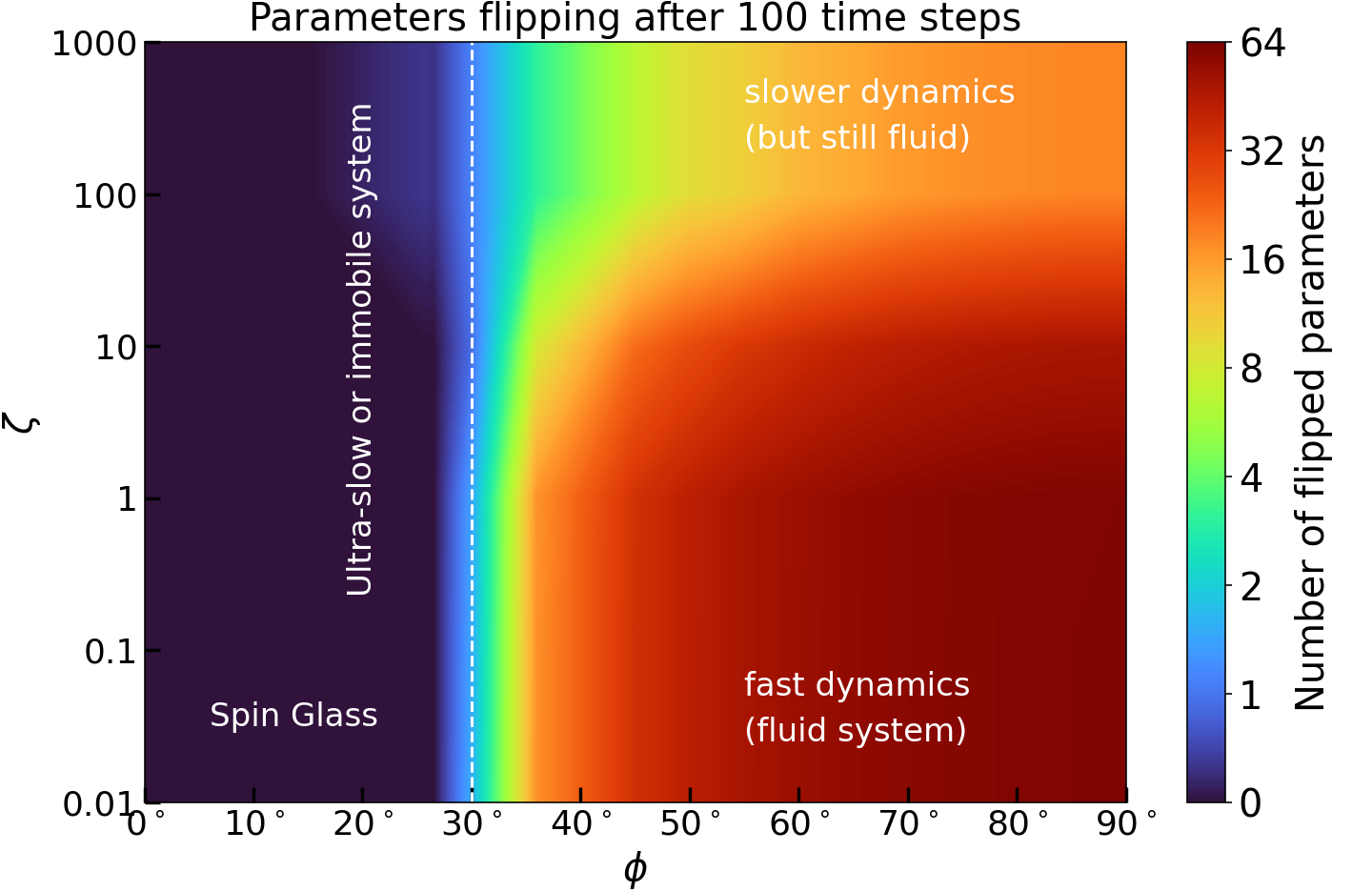}
        \caption{Phase diagram showing the number of parameter flips in the binary model of \cite{yarahmadi2026modeling}.}
        \label{fig:binary_model}
    \end{subfigure}

    \vspace{0.5cm}
    
    \begin{subfigure}{8.5cm}
        \centering
        \includegraphics[width=\textwidth]{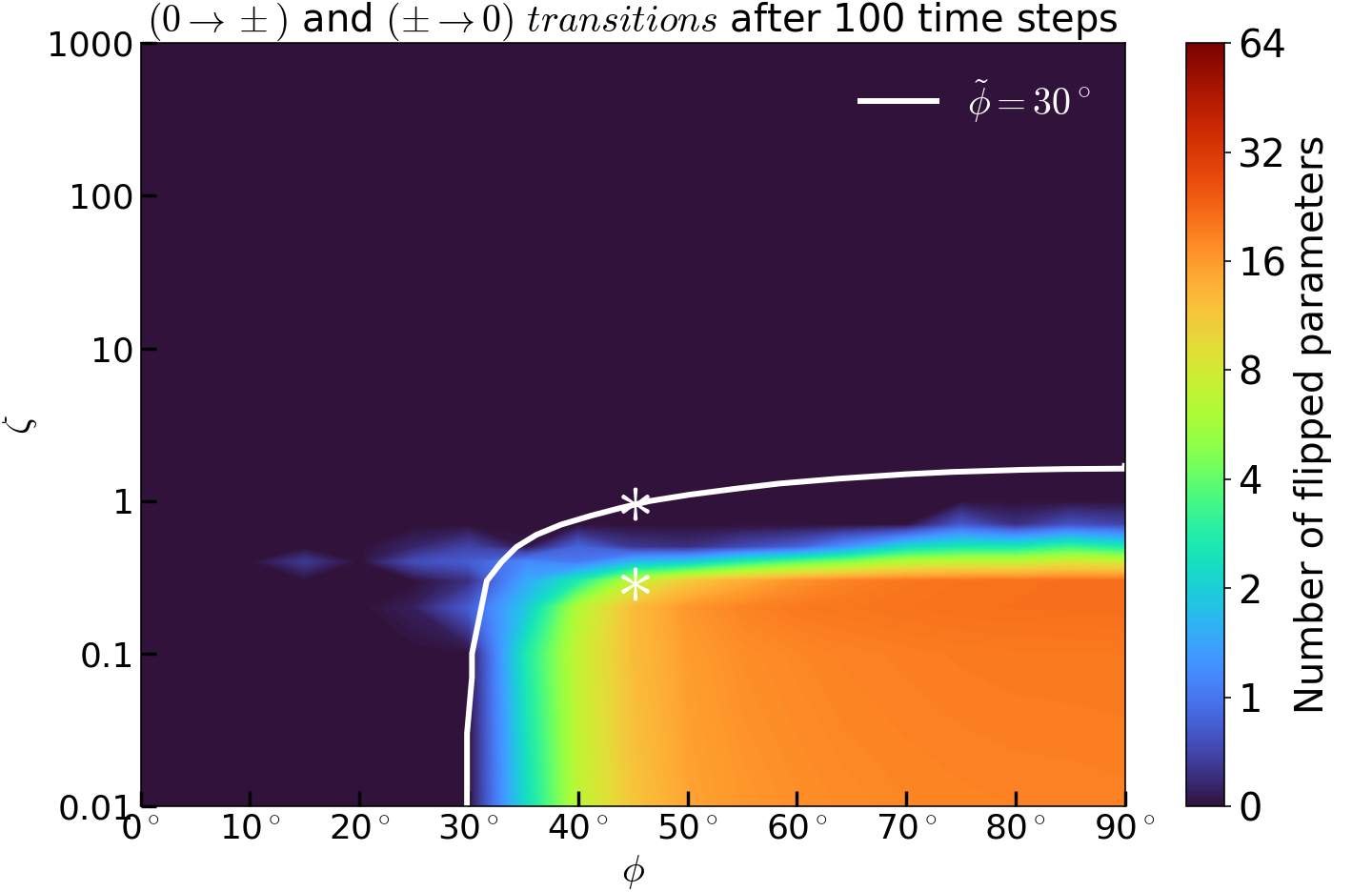}
        \caption{The number of parameters that enter or leave the 0 state at $t=100$, in the current model.}
    \end{subfigure}

    \vspace{0.5cm}

    \begin{subfigure}{8.5cm}
        \centering
        \includegraphics[width=\textwidth]{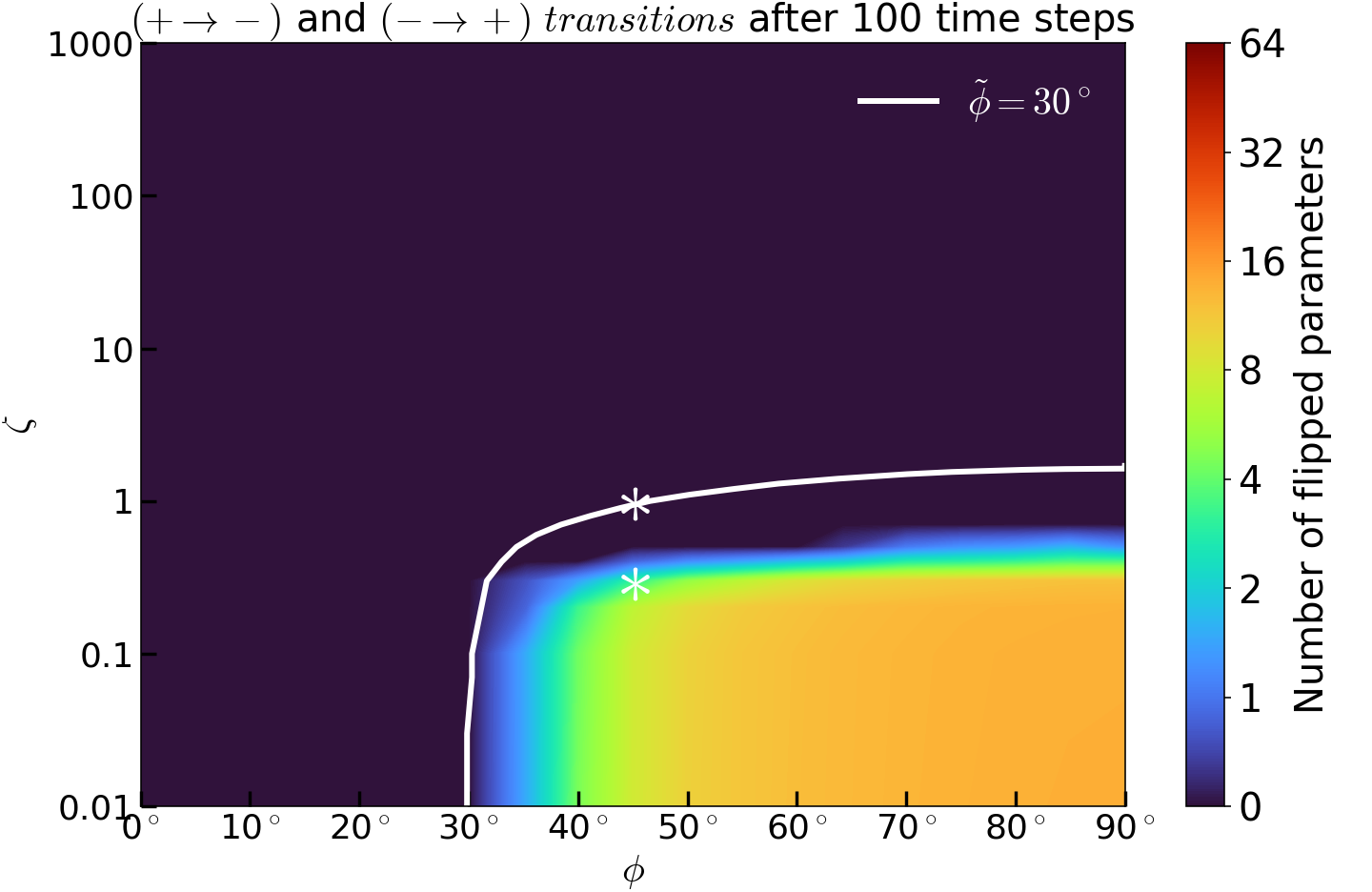}
        \caption{The number of parameters that flip between the values $\pm1$ at $t=100$, in the current model.}
    \end{subfigure}

    \caption{The heatmaps show the mean number of parameters flipping at the 100th time step, using a logarithmic color scale. The angle $\phi$ (in degrees) is on the x-axis, and $\zeta$ on the y-axis, on a logarithmic scale. The white curve shows $\tilde{\phi} (\phi,\zeta)=30^{\circ}$, while the white asterisks indicate the pairs of reference values ($\zeta=0.3,\;1$ and $\phi=45^{\circ}$) used in Figs.\ref{fig:58languages} and \ref{fig:distance_SD}.}
    \label{fig:fig3}
\end{figure}

\begin{figure}[htbp]
    \centering

    \begin{subfigure}{1\textwidth}
        \centering
        \includegraphics[width=\textwidth]{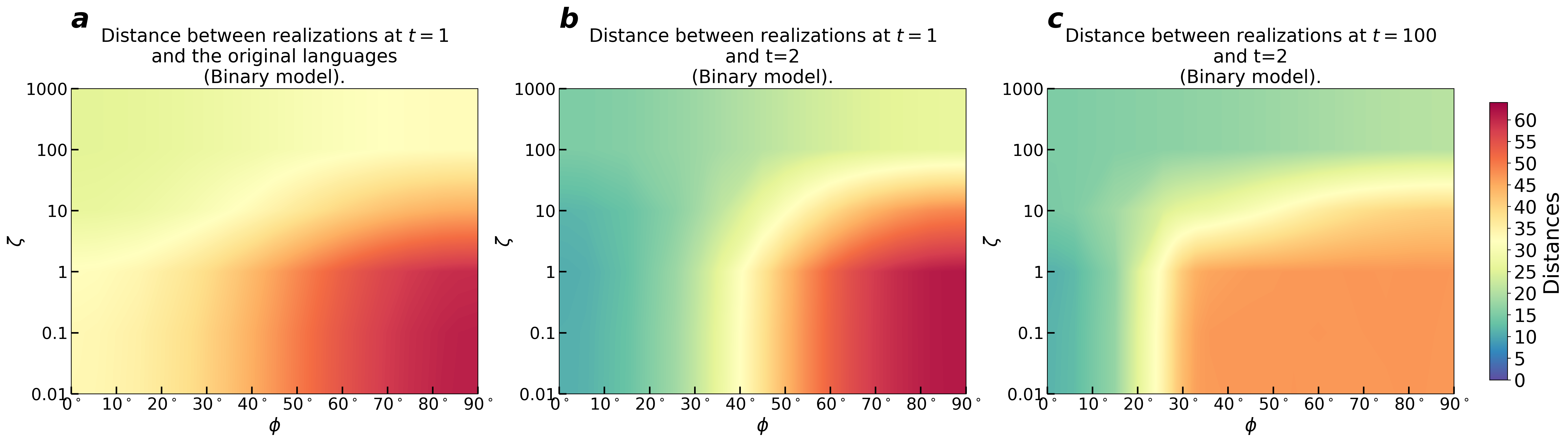}
        \caption{Binary model}
    \end{subfigure}

    \vspace{0.5cm}

    \begin{subfigure}{1\textwidth}
        \centering
        \includegraphics[width=\textwidth]{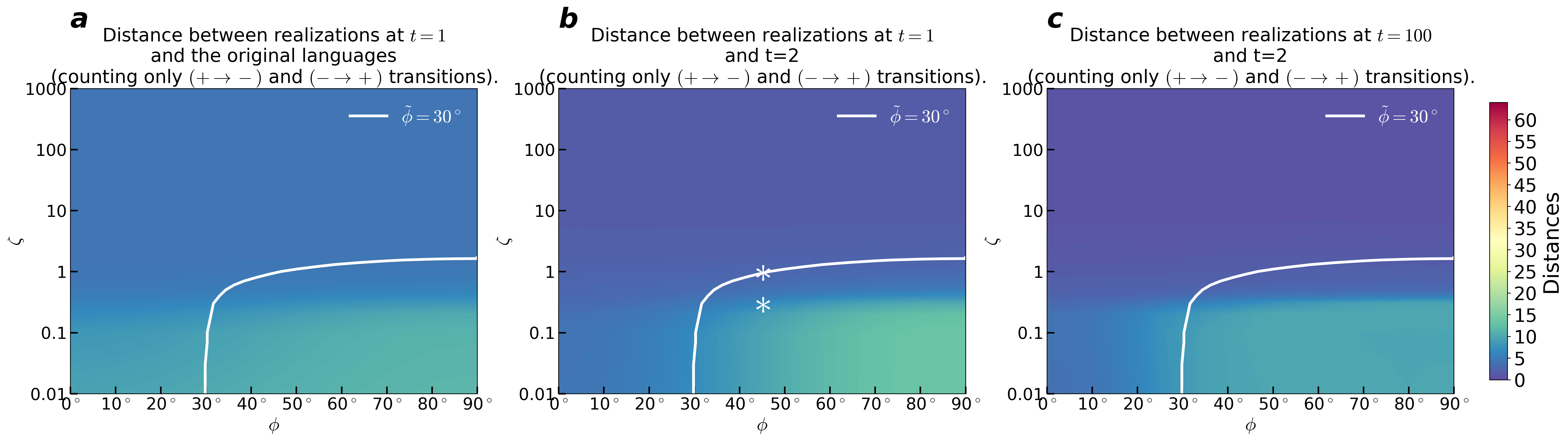}
        \caption{Counting only $(+\rightarrow -)$ and $(-\rightarrow +)$ transitions}
    \end{subfigure}

    \vspace{0.5cm}

    \begin{subfigure}{1\textwidth}
        \centering
        \includegraphics[width=\textwidth]{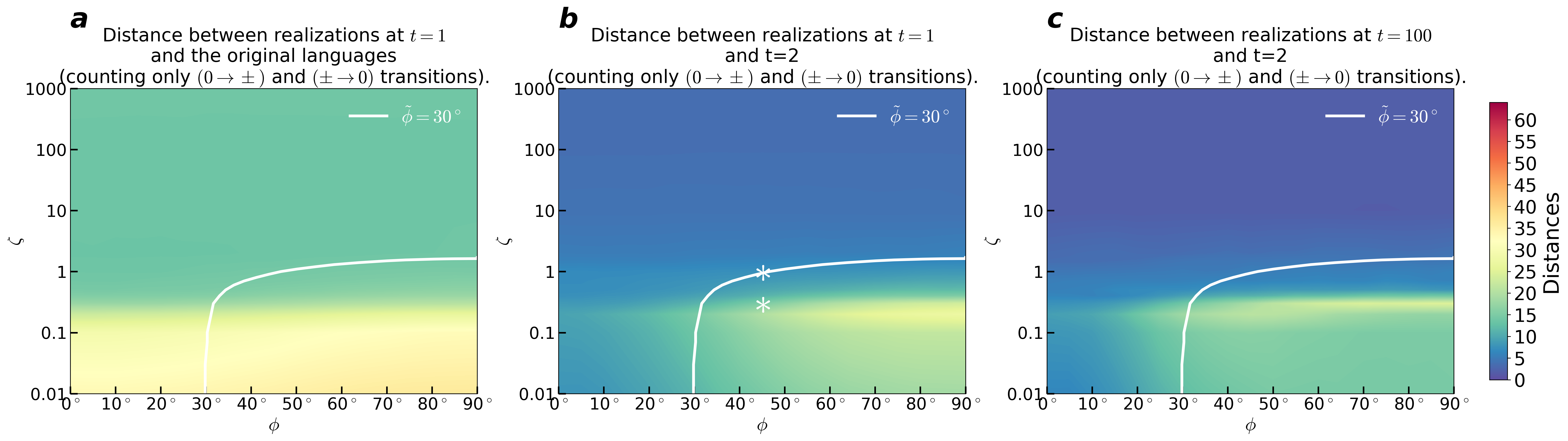}
        \caption{Counting only $(0\rightarrow \pm)$ and $(\pm\rightarrow0)$ transitions}
    \end{subfigure}

    \caption{The distance between the parameter vectors at $t=1$ from their original language values at $t=0$ (left column); the distance at $t=2$ from $t=1$, covered in the second time step (middle column); and the distance at $t=100$ from $t=2$ (right); showing that in the current model, even when the system of parameters remains fluid, the largest changes occur in the very first time steps. Note the linear scale of the heatmaps, unlike the log scale in Fig.\ref{fig:fig3}, while the white asterisks denote the same reference values.}
    \label{fig:distance_all}
\end{figure}

We also plot the phase diagram in Fig.\ref{fig:fig3} for the transitions to/from the zero states and between the nonzero values, and compare it with the one for the binary model in \cite{yarahmadi2026modeling}. When $\zeta >1$, the system gets stuck even at high values of $\phi$, as the overall coupling remains symmetric, but even more because the inferred weights dominate over the residual ones. As with the binary model, most of the dynamics unfolds in the first couple of time steps, which is shown in Fig.\ref{fig:distance_all} by plotting the average distance between the configurations at $t=0,1,2,100$. In the current model, most of the early transitions are to/from the undefined state -- they clearly reflect flips in the free parameters, that lead to readjustments of their implied ones; and also, as noted below, there are 4 types of transitions involving the 0-state, {\em vs.} only 2 skipping it. In the glassy regime, there is hardly any parameter changing value after 2 time steps, which in physical terms could be described as the temporal correlation length having become virtually infinite. Does it mean that the network of parameters keeps a memory of its original state? For that to happen, when it stops evolving, it should not always freeze into the same final state, a single attractor, but rather in a multiplicity of steady or quasi-steady states that reflects the initial starting points of the dynamics -- in our simulations, the 58 natural languages.

\subsection{Multiplicity of steady states}

\begin{figure}
        \centering
        \includegraphics[width=15cm]{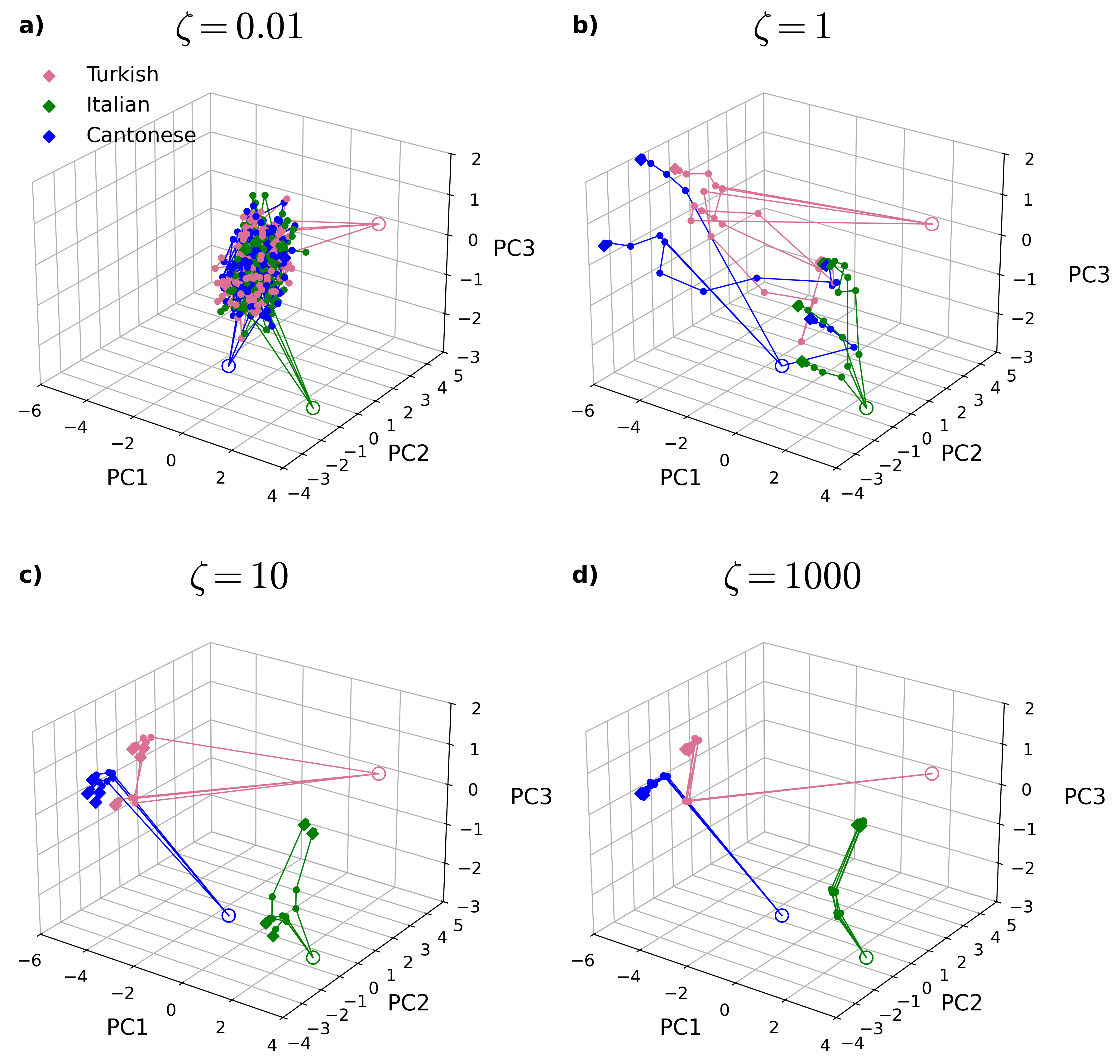}
    \caption{Four sample trajectories, each with a different assignment of residual weights, over 30 time steps starting from three syntactically distant languages (empty circles), as a function of $\zeta$, for $\phi=45^{\circ}$. The trajectories are illustrated in the space of the 3 Principal Components, extracted from the statistics of all 58 natural languages. For small $\zeta$ the system is fluid and rapidly loses memory of its initial condition, while for $\zeta \gtrsim 1$ the three languages tend to remain apart. For large $\zeta$ the residual weights become of course irrelevant, and trajectories essentially depend only on initial conditions, despite the weak variability induced by the random updating order.}
    \label{fig:4sets}
\end{figure}

Even within the glassy regime, where the isolated network sooner rather than later gets trapped in a steady state, the multiplicity of such steady states can vary a lot. This is shown in Fig.\ref{fig:4sets}.

For this and most other figures, we set $\phi=45^{\circ}$, to express a lack of assumptions about the symmetry or antisymmetry of the residual components of the weights. With this value of $\phi$, the system shows fluid dynamics for $\zeta \lesssim 0.4$ (see Fig.\ref{fig:fig2}), and definitely for $\zeta=0.01$ -- in which case, it should be noted in Fig.\ref{fig:4sets}a, the parameters keep evolving but within a restricted region. For $\zeta \geq 1$, instead, parameter values do not change further after the first few time steps, in any of the 12 sample trajectories. For very large $\zeta$ (Fig.\ref{fig:4sets}d), clearly the residual components of the weights, of minimal relative amplitude, do not matter, and the (brief) trajectories, unfolding just for 3 time steps, appear to depend on the starting points only, in this figure the 3 sample languages. Is this, then, a plausible regime?

\begin{figure}
        \centering
        \includegraphics[width=15cm]{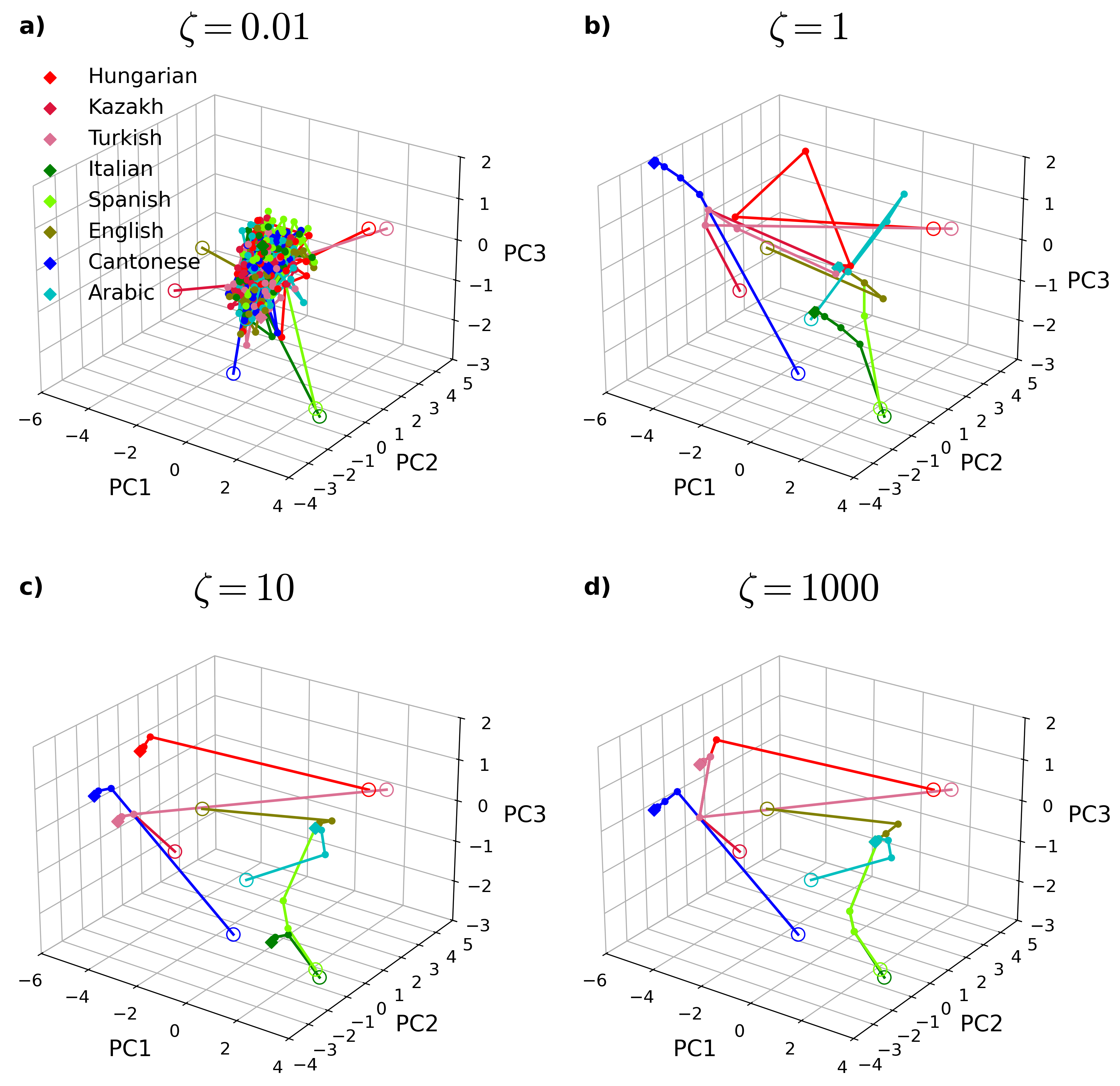}
    \caption{Sample trajectories starting from 8 different languages (the empty circles), all with the same assignment of residual weights, again shown up to $t=30$ in the space spanned by the first 3 Principal Components, as a function of $\zeta$, for $\phi=45^{\circ}$. Purely for illustration, we have chosen 3 Indo-European languages (2 of them Romance), 3 of the putative `Uralo-Altaic' group, and 2 thought not to be related to either superfamily, Cantonese and Arabic. While for small and intermediate values of $\zeta$ the figure appears to confirm the characterization expressed in Fig.\ref{fig:4sets}, for large $\zeta$ it indicates that without a significant contribution of the residual weights, different languages from the same family or superfamily eventually collapse onto each other, syntactically.}
    \label{fig:8languages}
\end{figure}

Fig.\ref{fig:8languages} indicates that this is not always the case, as it illustrates a non-trivial effect of the residual weights: even when they are the same across languages, they can still serve to keep {\em related} languages apart. This is suggested by Fig.\ref{fig:8languages}d, where, with $\zeta=1000$ and therefore negligible residual weights, after a few time steps, Hungarian, Kazakh and Turkish, i.e. the 3 `Uralo-Altaic' languages (a superfamily whose validity is still debated \cite{ceolin2021boundaries}) collapse onto each other; the same happens to the 3 Indo-European languages, attracting in their fold even Arabic, whose relation with the Indo-European superfamily, if anything quite remote, is controversial; only Cantonese, in this sample, remains distinct. A similar albeit weaker trend is seen in Fig.\ref{fig:8languages}c for $\zeta=10$, with the near-convergence of English, Spanish and Arabic (not Italian), and of Turkish and Kazakh (not Hungarian). For $\zeta=1$, instead, the trajectories are more diverse, and depend more substantially on the random assignment of residual weights. Therefore, vanishing residual weights do not support a sufficient multiplicity of attractor states. Even though network dynamics could be characterized as glassy, it is apparently not disordered enough to ensure that languages maintain their identity and potentially diversify.

\begin{figure}
        \centering
        \includegraphics[width=15cm]{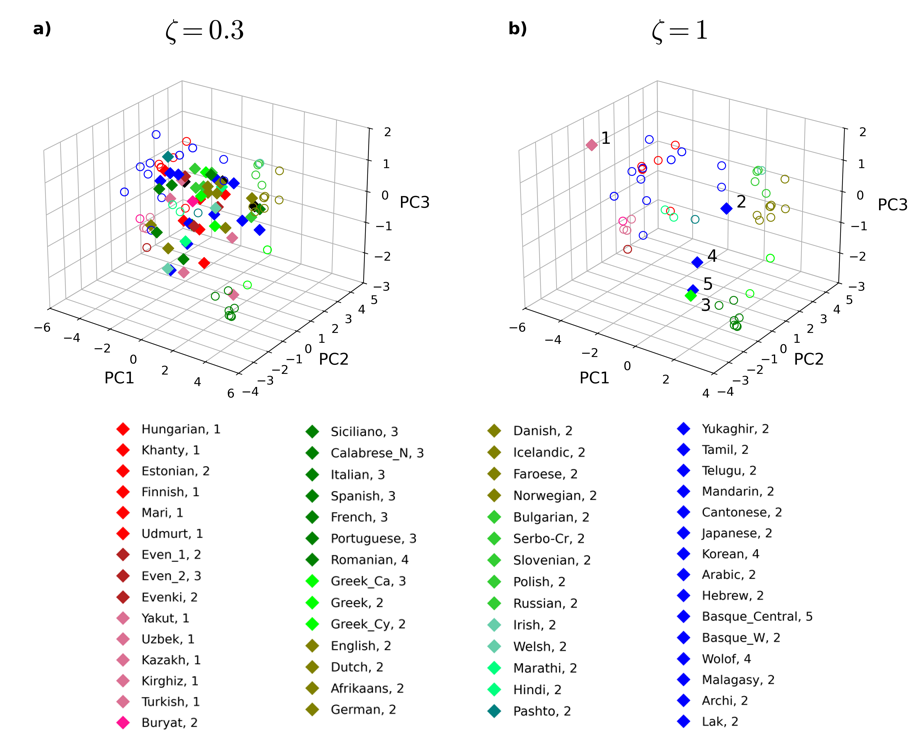}
    \caption{Sample, for one set of residual weights, of the initial (empty circles) and $t=30$ end points (full diamonds) of one trajectory for each of the 58 languages in the PCM database, in the space spanned by the first 3 Principal Components, for $\zeta=0.3$ (left) and $\zeta=1$ (right), and $\phi=45^{\circ}$. For $\zeta=1$ all 58 trajectories end up into just 5 distinct attractors, numbered in the right panel and referenced in the list of languages, which are color coded by language grouping.}
    \label{fig:58languages}
\end{figure}

In the following, therefore, we focus on $\zeta=0.3$ (in the fluid regime, but close to the glassy one) and $\zeta=1$ (which should be just within the glassy phase, in a very large system, but close to the fluid one), both with $\phi=45^{\circ}$, and we give a statistical look at the model dynamics for language families and for individual parameters.

We have color-coded the Indo-European family, which includes 28 languages in the sample, in shades of green corresponding to distinct sub-families; and partitioned the rest into an `Uralo-Altaic' superfamily, in shades of red, an `Other languages' group, in blue, with 15 languages each.
In the former, we have somewhat arbitrarily grouped 6 Uralic, 5 Turkic, 3 Tungusic and one Mongolian language, which were seen to cluster together to some extent, in \cite{ceolin2021boundaries} (see their Fig.1); while we have put all others, which appear to be unrelated or to show a significant relation with at most only another language, in the remaining group. 
Fig.\ref{fig:58languages}a shows that for $\zeta=0.3$ the model syntax of different languages would remain distinct, but lose memory of initial conditions -- the natural languages. The system is dominated by disorder and fluid, with no salient attractors, although there is a weak tendency for trajectories to be concentrated in a central region of PC space. Language families, color-coded in the figure, end up almost completely scattered after $t=30$ time steps.

For $\zeta=1$, instead, Fig.\ref{fig:58languages}b presents a completely different scenario: dynamical trajectories have collapsed onto just 5 attractor states, widely scattered and unevenly populated. Attractor state 1 (the numbering is arbitrary) collects the syntactic trajectories of 10 out of 15 `Uralo-Altaic' languages. The straddlers, however, apart from Estonian, are the 3 Tungusic and one Mongolic (Buryat) language in the PCM sample - the relation of these languages with the Turkic ones, once considered plausible, is now disputed \cite{pellard2022mirage}. All Turkic and all but one of the Uralic languages in the sample converge, therefore, on attractor 1, with this set of residual weights. Attractor 3 is the endpoint of 6 out of 7 Romance languages (except Romanian) plus Greek from Calabria (which might conceivably be attracted by the surrounding romance environment) and the Sebyan-Kyuyol variant of the Tungusic Even, likely a random effect. Attractor 4 collects three unrelated languages: Romanian, Korean and Wolof, and attractor 5, very close to attractor 3, only Central Basque. All other 36 languages converge onto attractor 2; they are mostly Indo-European, but that may be just because the IE family is over-represented in the PCM sample.

The exact pattern observed in Fig.\ref{fig:58languages} is specific to one random set of residual weights and one trajectory per language, but is representative of general features seen with other weight sets. It can be interpreted as the two $\zeta$ values chosen effectively bracketing a cross-over range. At either end of this range, syntax dynamics in the model are implausible, but somewhere within the range the system of syntactic parameters, left to its own interactions, may show a tendency to find one of several quasi-stable states, to some degree correlated with initial conditions.

\subsection{Variability within and across languages}

\begin{figure}
\centering
         \includegraphics[width=15cm]{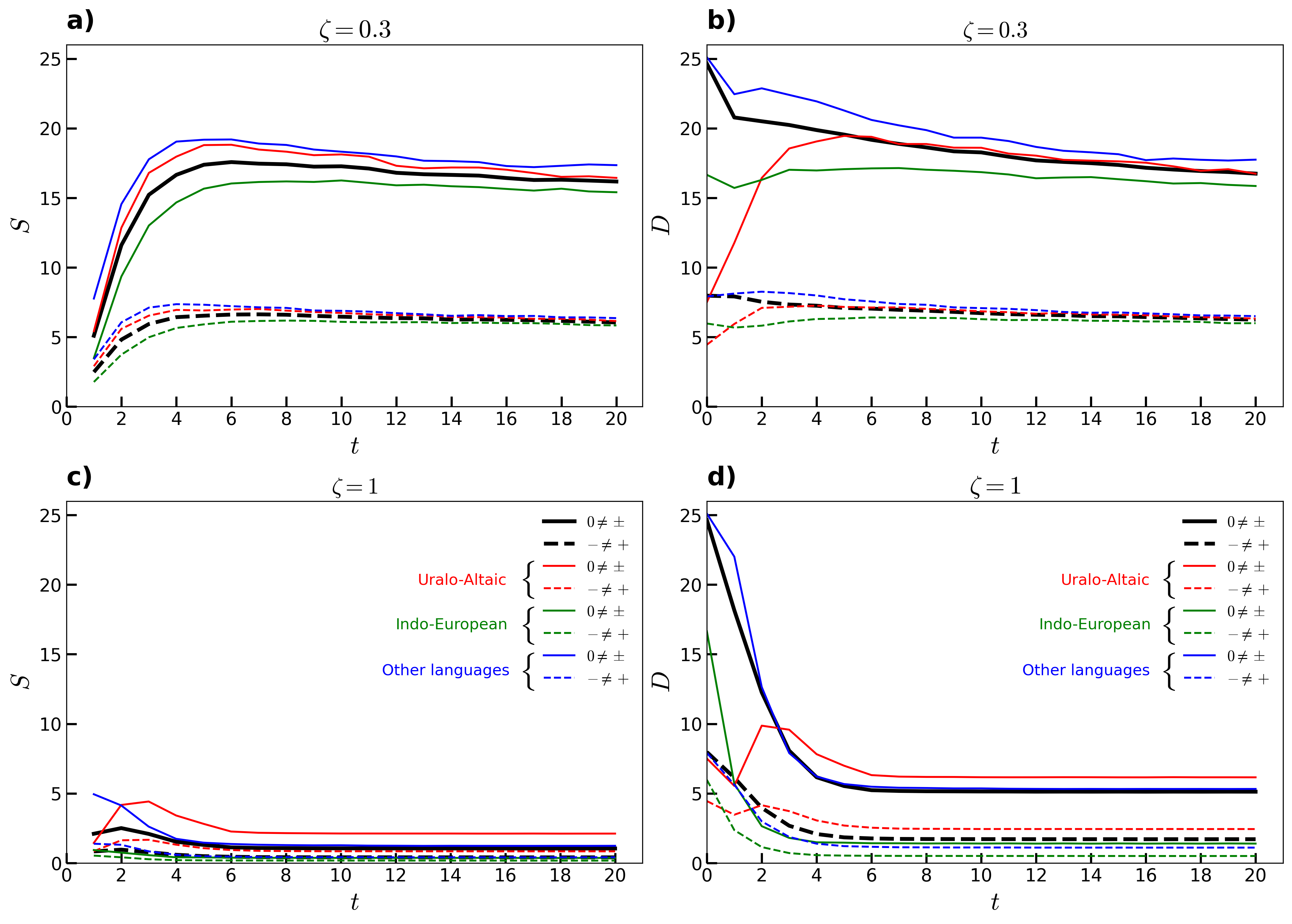}
        \caption{The time course of the average distance, on the left, between evolutionary trajectories starting from the {\em S}ame original language and subject to the same interaction weights (the trajectories differ only because of the randomized updating of the parameters at each level); and, on the right, of the average distance between evolutionary trajectories starting from 
        {\em D}ifferent original languages. For $\zeta=0.3$, above, and $\zeta=1$, below, both with $\phi = 45^{\circ}$. Both measures are summed over parameters and averaged over languages (or language pairs in the right panels).}
    \label{fig:distance_SD}
\end{figure}

These insights are supported by Fig.\ref{fig:distance_SD}, which shows the distance, in terms of the number of parameters with different values, within trajectories starting from the same language ($S$) and between languages ($D$). Both measures are obtained with a single set of weights, with $N_r=5$ runs, and then are averaged over $N_w=50$ sets of residual weights. The differences involving the 0 state are counted separately from the $\pm 1$ differences.
In general, differences involving the 0 state tend to be doubly as numerous as those not involving it, which is expected considering that 2 configurations can differ in one parameter in 4 different ways in the former case (0/+1, 0/-1, +1/0, -1/0), and only in 2 in the latter (+1/-1, -1/+1). 

For $\zeta=0.3, \phi=45^{\circ}$, the system is fluid, and the trajectories quickly lose memory of their initial condition (Fig.\ref{fig:distance_SD}, above, see how the {\em S} measure approaches the values of the {\em D} measure). The only significant differences are in the first few time steps, when the {\em S} measure grows, starting from 0 at $t=0$, while the {\em D}  measure starts from the values taken across natural languages, and overall tends to decrease -- except, initially for the usual suspect, the `Uralo-Altaic' putative superfamily.

With $\zeta=1$, instead (still $\phi=45^{\circ}$) the lower panels of Fig.\ref{fig:distance_SD}, show that the overall trend, across all languages and language pairs (thick black curves), is for the distances to decrease and then stabilize without further changes after 7-8 time steps. Across all languages, distances between pairs tend to be 4-5 times larger than between different trajectories from the same starting languages (the same scale is used on all {\em y}-axes). Thus, the syntactic diversity currently expressed across natural languages, at least for those in the PCM sample, would be somewhat maintained, although much reduced, in the model trajectories run with the reference values $\zeta=1, \phi=45^{\circ}$. 

These overall trends are strongly affected, however, by the relations among the 58 languages represented in the PCM database. The Indo-European family including 28 languages shows substantial convergence. The average syntactic distance between Indo-European languages, about 23 parameters at $t=0$ (summing the two green curves in the right lower panel, Fig.\ref{fig:distance_SD}b) fizzles out for $\zeta=1$ to about 3 after only 4 time steps, and ca. $25\%$ of it is due to intrinsic variability among different trajectories starting from the same language (left lower panel). This is in contrast with both the `Uralo-Altaic' and the `Other languages' groupings, each including 15 members each. Interestingly, while the `Other languages' show trends quantitatively similar to those seen across all languages, the putative `Uralo-Altaic' group stands out, by presenting both between languages and within language variability above the grand averages (still for $\zeta=1$). The within language variability ($S$) is roughly double the average value, for both types of transitions. Even more striking, their distances grow over the first 2-3 time steps (see the red curves in the lower panels), to stabilize at lower values later on. It is as if there is something intrinsically `unsettled' in the syntactic parameters of the Uralo-Altaic languages, which may have to do with the PCM database being dominated by the Indo-European languages. A similar effect is also seen for $\zeta=0.3$ (the {\em D} measure in the upper right panel, red curve).

\subsection{Robust and fragile parameters}

Which parameters are more exposed to change, in the model? Since most of the changes in their value occur in the first few time steps, largely in the very first, we focus here on what happens between $t=0$ (the syntax of the original languages) and $t=1$.
We lump together all 6 types of transitions.

\begin{figure}
        \centering
        \includegraphics[width=13cm]{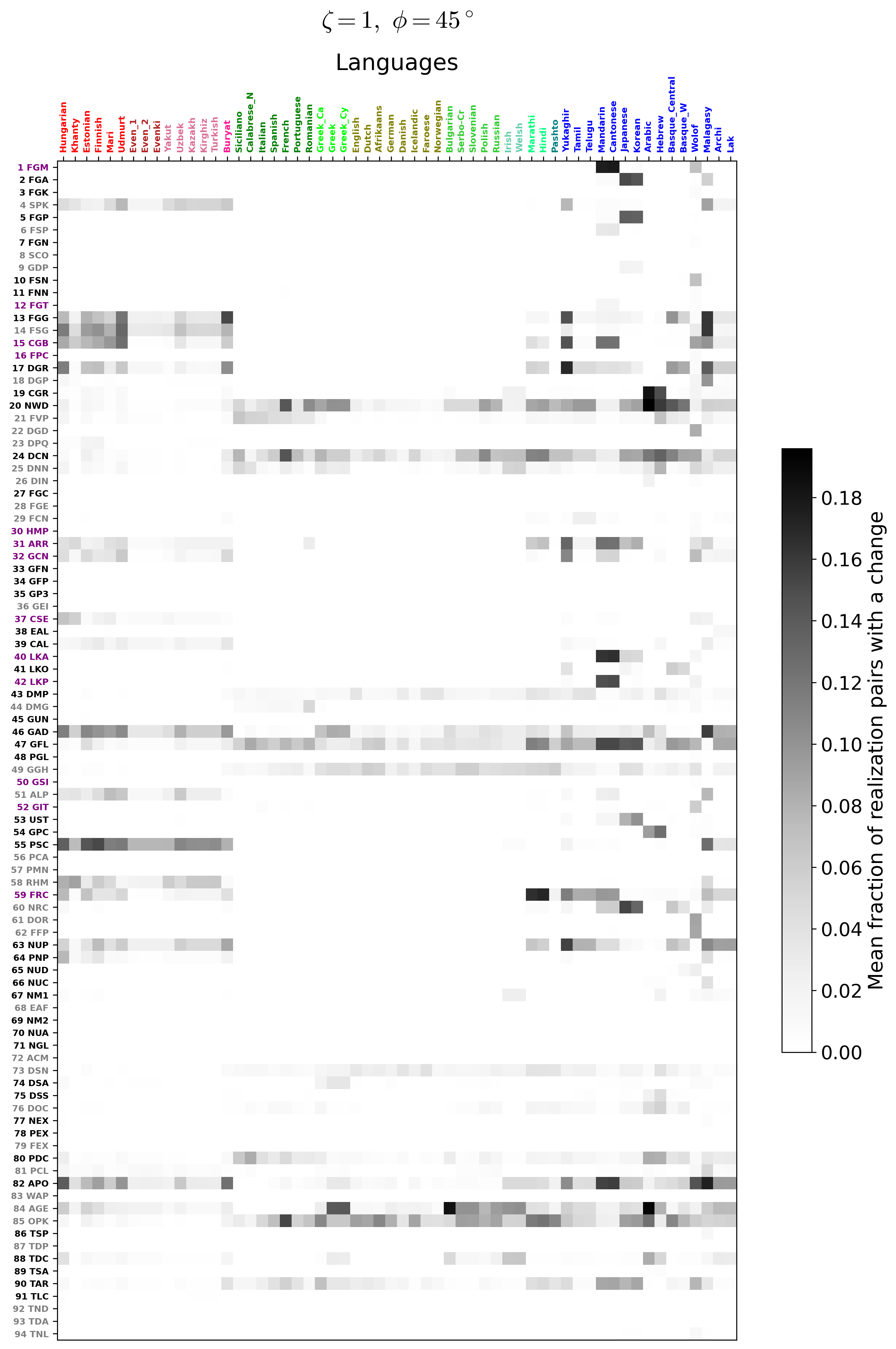}
        \caption{The fraction of occurrences of a change in value of each parameter in the trajectories starting from each original language, in the first time step, averaged not only over $N_r=250$ runs per language but also $N_w=250$ different sets of weights.}
    \label{fig:av_heatmap_first_step}
\end{figure}

As shown in the two examples provided in App. \ref{app:fragile}, for each set of residual weights there is a relatively sparse sample of parameters, scattered among languages but particularly concentrated in some `brittle' languages, that tend to change value during the first time step. Whether they do or not depends on the remaining source of randomness, the updating order. This fragile sample depends on the set of residual weights, but it is still not equidistributed across parameters and languages, as shown by averaging over $N_w=250$ different weight sets, in Fig.\ref{fig:av_heatmap_first_step}. Value changes during the first step remain unevenly distributed across languages (along the {\em x}-axis) and parameters (along the {\em y}-axis). 

Predictably, the languages which overall seem more prone to syntactic change are those in the `Other languages' group, followed by those in the `Uralo-Altaic' group, particularly the Uralic ones and Buryat, the single Mongolic language included in the sample. This presumably reflects the Indo-European dominance of the PCM database (in its current form), and should not be taken to imply an intrinsic fragility of the languages themselves.

Potentially more informative is the variable susceptibility to change among the parameters. Here, however, we do no see any simple trend related to the position of the parameter in the mangrove of implications of Fig.\ref{fig:implications}: some of those prone to change value are free parameters, some are intermediate, and others are without dependents, at the bottom of the mangrove. This issue is left, therefore, for deeper linguistic-informed analyses.

\subsection{Robust and dubious interactions}

The model suggests the need for linguistic analyses even independently of its dynamical features, glassy or not; solely on the basis of which parameter appear to interact more than they would be expected on the basis of implications alone.

Consider any pair of parameters $i$ and $j$ that are defined in at least a subset of the 58 natural languages in the PCM database, and the empirical frequencies 
$f^{++}$, $f^{+-}$, $f^{-+}$, and $f^{--}$ corresponding to the occurrence of the four possible combinations of parameter values when both parameters are defined.

If the two parameters were statistically independent, these frequencies would approximately satisfy
\begin{equation}
f^{++} f^{--} \simeq f^{+-} f^{-+}.
\end{equation}
Any large deviation from this approximate equality may provide evidence for an interaction between the two parameters, and we have indeed quantified this deviation through the inferred interaction weight $w^I_{ij}$ defined in Eq.~\ref{w-no-alpha} (with the standard regularization applied to the frequencies $f^{++}$, $f^{+-}$, $f^{-+}$, and $f^{--}$ to estimate probabilities).

The inferred weights capture statistical dependencies between syntactic parameters. However, when two parameters are embedded within the same implication network, part of the observed interaction may arise trivially from their hierarchical relationship. 
In particular, we can see in Fig.\ref{fig:implications} that some parameters (e.g., 1, 5, 7, 10, 17, ...) appear to act as sort of \textit{syntactic hubs}. These parameters are above many other parameters in the implicational hierarchy and therefore exert an indirect influence on a large portion of the network. Consequently, two parameters may appear to have a strong inferred interaction simply because they share many common ancestors, rather than because they are directly related.

\begin{figure}
\centering
\includegraphics[width=0.48\linewidth]{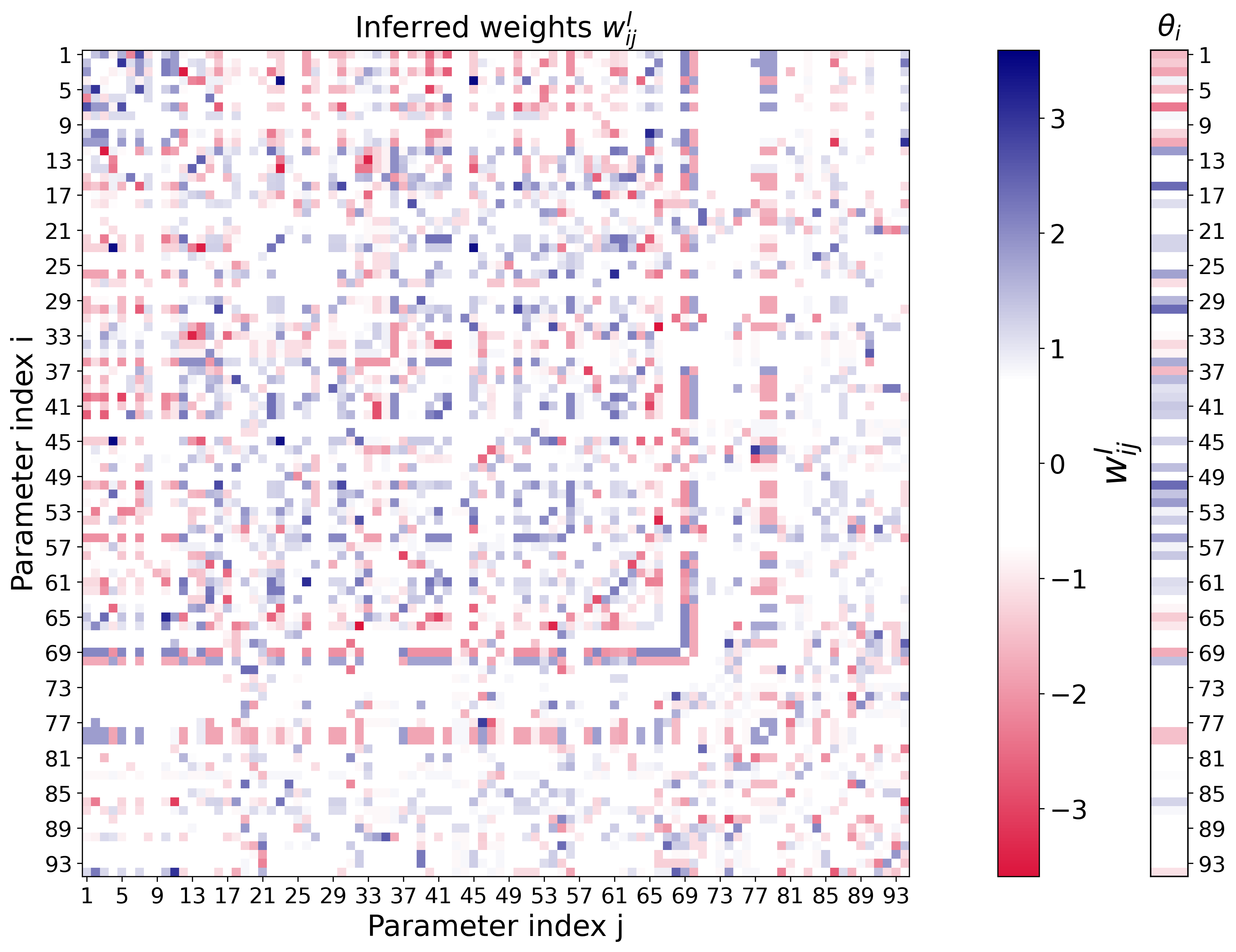}
\hfill
\includegraphics[width=0.47\linewidth]{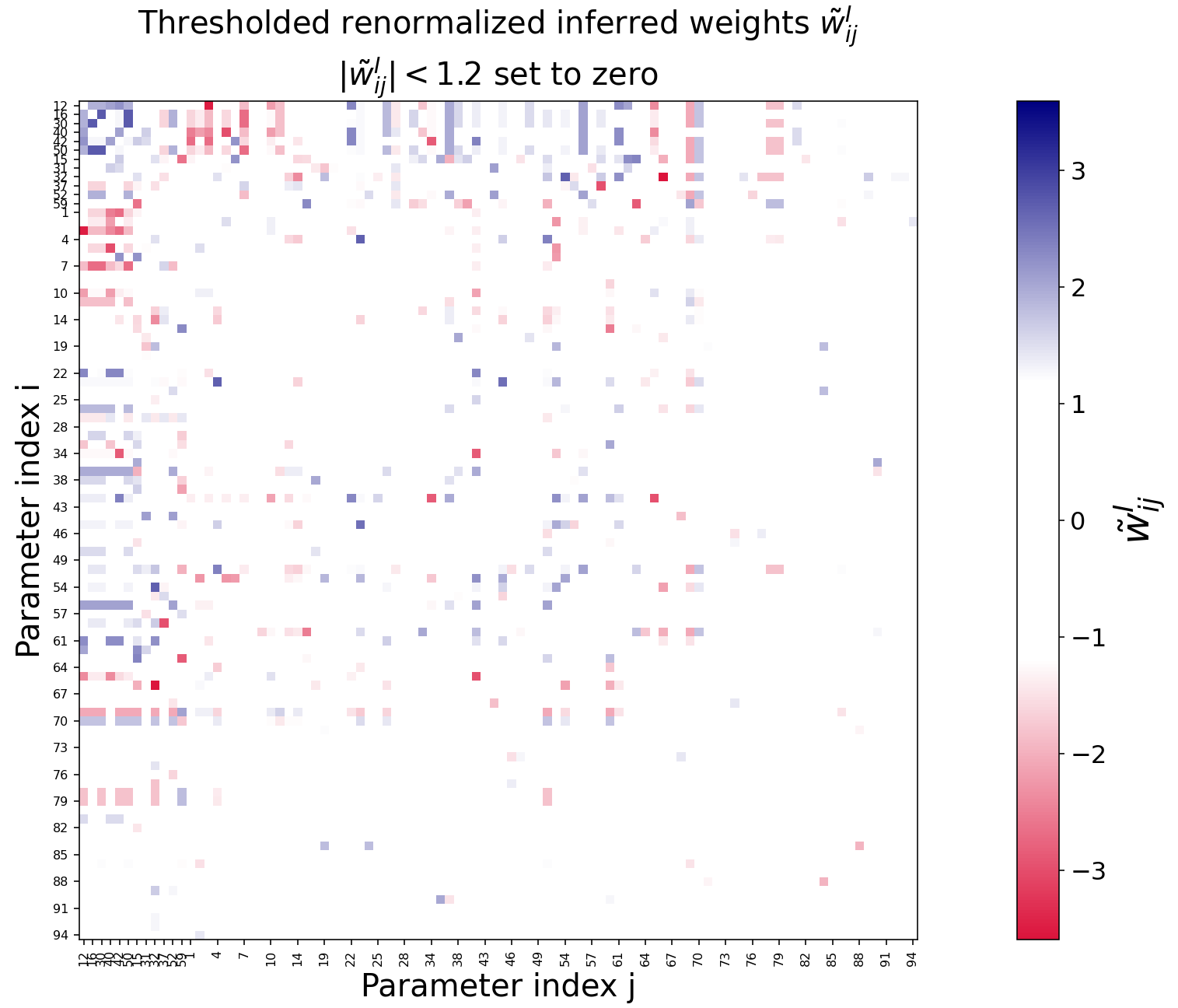}
\caption{Heatmap of the inferred weights, in their original values (left) and after the renormalization intended to suppress the effects of implications, thresholding at $|w^I_{ij}| > 1.2$, and sorting the parameters, in particular the free ones, for ease of visualization (right).}
\label{fig:fig7}
\end{figure}

To distinguish direct interactions from effects induced by the structure of implications, we introduce an {\em ad hoc} renormalization procedure for the inferred weights, described in App.\ref{app:renorm}. 
Further, to reduce the visual contribution of noise, the inferred weights with magnitude smaller than a threshold are removed, i.e., set to zero, in Fig.\ref{fig:fig7}(right)
\begin{equation}
\tilde{w}^I_{ij}=0 \qquad \text{for} \qquad |\tilde{w}^I_{ij}|<1.2.
\end{equation}

Finally, we sort the rows and columns in the matrix on the right of Fig.\ref{fig:fig7}, to make more salient the interactions, particularly among the free parameters.
 
It makes sense to focus on the subset of 13 free parameters, as they are defined independently in all languages and are not constrained by implication relations. Therefore, any interaction inferred between these parameters cannot be attributed to shared hierarchical dependencies, but instead reflects either genuine coordination or statistical fluctuations.
The resulting interaction matrix appears to show a striking organization among the free parameters. In particular, parameters 12, 16, 30, 40, 42, and 50 form a coherent interaction cluster. Since these parameters are free from implication constraints, their interactions survive the renormalization procedure.

Several of the parameters in this cluster show strong positive mutual interactions, and tendentially negative ones with the large implication hierarchy descending from parameter 1, including parameters 3, 5, 7, and 10.
Thus, although these parameters are not connected through implication relations, they appear to behave collectively as a dynamical subsystem within the syntactic parameter space.
This behavior is not shared by all free parameters. The remaining free parameters (15, 31, 32, 37, 52, and 59) display a different pattern, characterized by weaker and more scattered interactions. 
Among the dependent parameters, there are some (e.g., 22, 26, 36, 56, 61, and 62) that seem to interact positively with the first cluster, despite the renormalization. Interestingly, they do not possess implication descendants of their own. Therefore, they appear to be structurally embedded within the main hierarchy while being interactionally aligned with that cluster.

It should be noted, though, that this clustering might reflect primarily the extremely rare occurrence of some parameter values in the PC database. Thus, parameters 16, 30 and 50 take only the -1 value in the PCM database (they take the value +1 in other language currently not included in the database) and the interactions we might infer between them and other parameters could be merely an artifact.  
Because of this, we leave to future studies, which may benefit from a database expanded both in terms of parameters and of languages, a more linguistically oriented analysis of the inferred interactions. 

\section{Discussion}

Languages evolve, of course, subject to a vast array of socio-historical factors of disparate nature and unfeasible quantitation, which a simplified mathematical model can hardly contemplate to describe. What we can hope to get insights on are general constraints on syntax dynamics due to internal factors, where `internal' refers to the language system and to how it is acquired and implemented in the brain of speakers. In this respect, the Principles and Parameters formulation appears to be particularly useful, as it facilitates the mapping of a closed system of interacting parameters onto a closed system of interacting spins, of the type much studied in statistical physics. Three main limitations restrict the validity of this mapping, beyond the unavoidable focus on an isolated system: the implications, a peculiar set of linguistic constraints that make parameters unequal and forces some of them to be, at certain times, in a `0-state'; the small number of parameters for which data is available, which in the case of the PCM database also come solely from an analysis of the nominal phrase, not of other components of sentence structure; and the limited amount of `experimental data', in the case again of the PCM database the 58 languages it currently includes, disproportionately biased towards Eurasian language families. 

Within such limitations, the model shows the signature, made somewhat opaque by heterogeneity and finite size, of a large multi-dimensional dynamical system approaching the chaotic regime -- or rather, {\em exiting} the chaotic regime as $\zeta$ grows. In the chaotic regime, for $\zeta \ll 1$, the residual weights dominate over the inferred ones. As they are assumed to be asymmetric, $\phi=45^{\circ}$, the compound asymmetry angle $\tilde{\phi}$ of the overall weights $\{J_{ij}\}$ is above the critical angle $30^{\circ}$ that signal the transition from stable to fluid parameter configurations \cite{yarahmadi2026modeling}. Fluid parameter configurations do not seem to be compatible with the notion of a stable syntax, that has reached, in each particular language, an approximate attractor or metastable state \cite{amit1989modeling}. 

For $\zeta > 1$, instead, the inferred weights, symmetric by construction, dominate over the residual ones and stabilize the parameters in a glassy phase, as parameters tend to quickly freeze in some configuration, but in general a frustrated configuration, far from compatible with each and all the biases among the parameters. The multiplicity of such configurations is limited, however, and does not seem to be able to span the diversity observed in natural languages. 

A plausible scenario, which combines stability with multiplicity, may emerge in the intermediate range, roughly $0.3 < \zeta < 1$ in our simplistic model, a range with similarities to the so-called critical region of a very large system of interacting units near a phase transition. Criticality has been argued to be a key feature of brain function, although its exact nature and role have been interpreted in a variety of ways \cite{tomen2019functional}. From the point of view of the long-term dynamics of syntactic parameters, in the critical region temporal autocorrelationss are expected to extend over long time scales, allowing for syntax stability, while remaining sufficiently close to the fluid phase to allow for language change. Such proximity may also be essential for the ability to acquire multiple languages. 

Beyond these general considerations, the model may afford some specific insights on specific parameters and their implementation in the brain. First, we have seen that even after averaging over many sets of residual weights, the propensity for change remains unevenly distributed across parameters and across languages. In some way, this should reflect a measure of non-adaptation to the implicational structure of the parameters, and to the prevailing patterns that such structure expresses in natural languages. It would be premature to draw more concrete conclusions or predictions at this stage, but this may be possible with a suitably extended database.

Second, particularly strong couplings between specific pairs of parameters, which appear to characterize a sparse set among all possible pairs of parameters, might point to how they actually interact in the brain. The study of how aspects of syntax are expressed in brain activity is still in its pre-infancy \cite{sauppe2021neural}, but it is likely to become  of focus of exciting research in the forthcoming years. Also in this respect, concrete predictions about neural interactions based on strong inferred weights should wait for the extension of the PCM database to more languages and more parameters. Nevertheless, the kind of analyses discussed here may contribute to elucidate that most intractable of neuroscience issues, how language is produced and understood by the brain. 

\subsection*{Acknowledgments}

Dr Kwang Il Ryom was involved in the early stages of this study, and his contribution is gratefully acknowledged, as are extensive discussions with Giuseppe Longobardi, Paola Crisma and Cristina Guardiano, who have over several years established the PCM database, as well as with Eric De Giuli. 

\section*{Appendices}

\appendix

\section{The effect of the regularization}
\label{app:regul}
\begin{figure}
\centering
\includegraphics[width=0.7\linewidth]{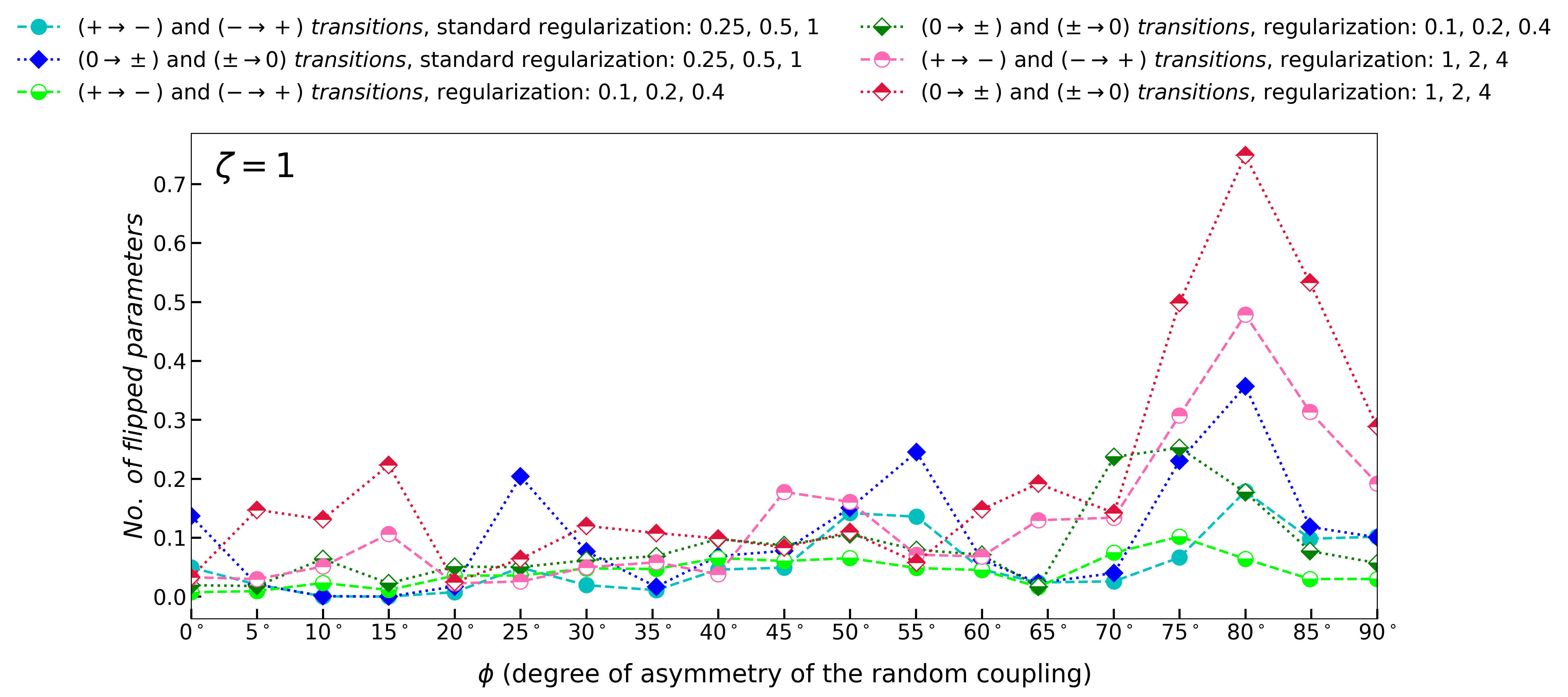}
\caption{The average number of parameters that change from defined to undefined or viceversa and of those that flip between their $\pm 1$ values, at time step 100, for $\zeta=1$ as a function of $\phi$ and of the regularization adopted (two variants of the standard regularization are shown, with the regularizing terms multiplied by 0.4 or by 4). Note the linear {\em y}-scale, which leads to seemingly larger finite-size fluctuations than those shown in Fig.\ref{fig:fig2}.}
\label{fig:fig2reg}
\end{figure}

The exact number of parameters still changing value at $t=100$ in the glassy regime is seen to depend moderately on the strength of the regularization used to estimate probabilities from frequencies, as shown in Fig.\ref{fig:fig2reg}. Based on this observation, we adopt the `standard' regularization detailed in the text for all the analyses.

\section{Quantifying the asymmetry of the full couplings}
\label{app:asymm}
It is easy to see that the degree of asymmetry of the residual weights, set {\em a priori} by the value of $\phi $, could be also estimated {\em a posteriori} by decomposing the coupling matrix into its symmetric and antisymmetric components, as
\[
\phi \simeq
\tan^{-1}
\left(
\frac{\left\|W^R_{ij}-W^R_{ji}\right\|_{F}}
{\left\|W^R_{ij}+W^R_{ji}\right\|_{F}}
\right),
\]
where $\|\cdot\|_{F}$ denotes the Frobenius norm, $\|W\|_{F}=\sqrt{\sum_{i,j}W_{ij}^{2}}$. The numerator measures the magnitude of the antisymmetric component of the coupling matrix, while the denominator quantifies the magnitude of the symmetric component. Therefore, $\phi$ provides a dimensionless measure of the departure from symmetric (reciprocal) coupling.

We can similarly decompose the full couplings $J_{ij}$ into symmetric and antisymmetric components, $J_{ij}=S_{ij}+A_{ij}$, where $S_{ij}=\frac{1}{2}(J_{ij}+J_{ji})$ and $A_{ij}=\frac{1}{2}(J_{ij}-J_{ji})$ and obtain a dimensionless measure of their deviation from symmetry as
\[
\tilde{\phi}=
\tan^{-1}
\left(
\frac{\|A\|_{F}}{\|S\|_{F}}
\right).
\]
The angle $\tilde{\phi}$ quantifies the relative strength of the antisymmetric contribution: $\tilde{\phi}=0$ corresponds to a purely symmetric coupling, $\tilde{\phi}=45^{\circ}=\pi/4$ to asymmetric couplings, whereas larger values of $\tilde{\phi}$ indicate increasingly antisymmetric interactions.

\section{Variability in parameter susceptibility}
\label{app:fragile}

\begin{figure}
        \centering
        \includegraphics[width=13cm]{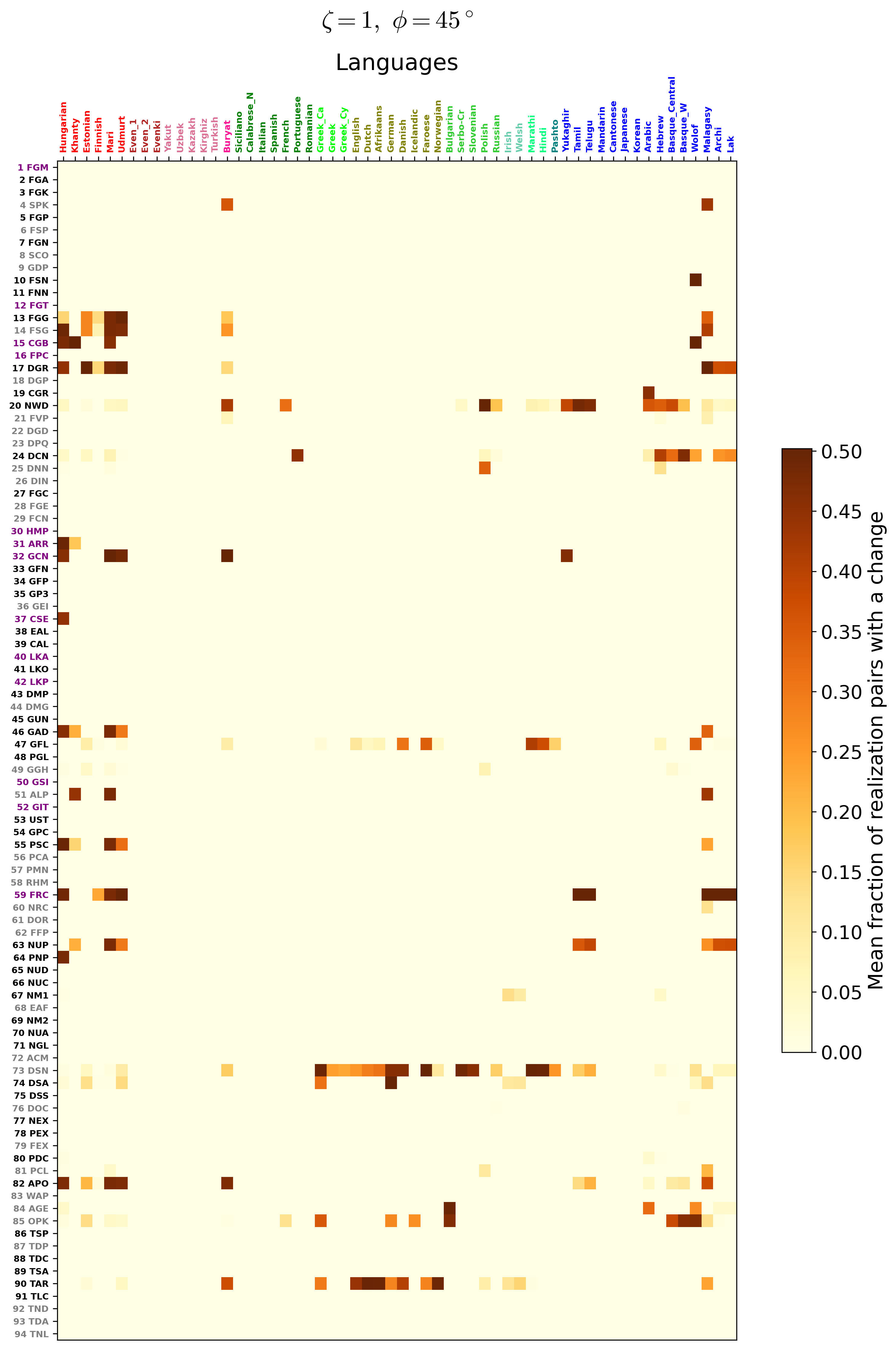}
\end{figure}
\begin{figure}
        \centering
       \includegraphics[width=13cm]{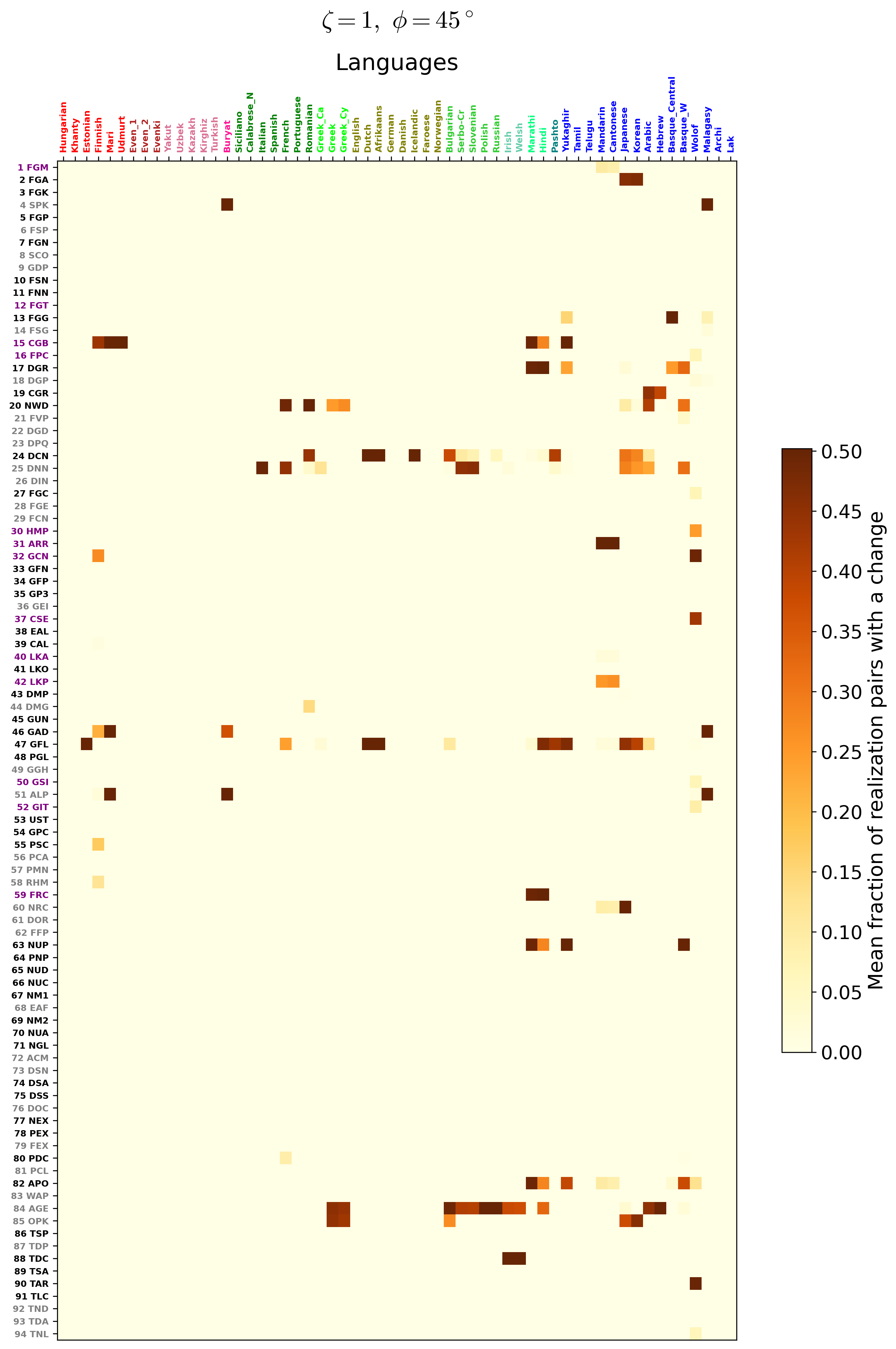}
        \caption{The fraction of occurrences of a change in value of each parameter in the trajectories starting from each original language, in the first time step. Two examples are given, each for a fixed random assignment of the residual weights and $N_r=250$ trajectories per language. Again, the only source of variability is the randomized updating of the parameters at each level $l$.}
    \label{fig:heatmap_first_step}
\end{figure}
The examples of Fig.\ref{fig:heatmap_first_step} show, for two sets of residual weights, that value changes during the first step are very unevenly distributed across languages (along the {\em x}-axis) and parameters (along the {\em y}-axis). 

\section{Renormalization procedure for the inferred weights}
\label{app:renorm}
The aim of this {\em ad hoc} procedure is to make salient the inferred couplings which are presumably strong and direct, suppressing those that may be largely cascaded indirectly through the implicational hierarchy.

To this end, we renormalize the implication contribution \(w^{I}_{ij}\) according to the amount of weighted ancestry shared by parameters \(i\) and \(j\). Specifically, we define the renormalized interaction as
\begin{equation}
\tilde{w}^{I}_{ij}
=
\frac{w^{I}_{ij}}{1+A_i^j+A_j^i},
\label{renorm_eq}
\end{equation}
where \(A_i^j\) and \(A_j^i\) quantify the weighted common ancestors of parameters \(i\) and \(j\), measured from the perspectives of \(j\) and \(i\), respectively. We assign each ancestor a weight that decreases by successive factors 1/2 with its distance from the parameter. If an ancestor is reached after \(n\) steps in the implication tree, its weight is defined as
\[
A_i(k)=\left(\frac{1}{2}\right)^n.
\]
Note that this is analogous to genetic inheritance: an individual inherits one half of their genes from each parent, one quarter from each grandparent, one eighth from each great-grandparent, and so on. Note also that \( 
A_i(i)=1.
\)
If the same ancestor can be reached through multiple implication paths, only the largest weight is retained, corresponding to the most immediate relationship (the shortest path in the ancestor tree). Therefore, each entry \(A_i(k)\) contains the maximum scaled contribution of parameter \(k\) as an ancestor of parameter \(i\).

Repeating this procedure for all 94 parameters produces the full \(94\times94\) matrix \(A_i(k)\). Each row \(i\) contains the scaled ancestor information of parameter \(i\), while each column \(k\) represents the contribution of parameter \(k\) as an ancestor.
Then the shared heritage between two parameters \(i\) and \(j\) is quantified through their shared weighted ancestors. For each pair of parameters \(i\) and \(j\), we sum the contributions of all ancestors \(k\) that are present in the ancestor list of \(j\):
\begin{equation}
A_i^j=\sum_k A_i(k) \quad \text{if } A_j(k)>0 .
\end{equation}

The resulting matrix \(A_i^j\) is another full \(94\times94\) matrix (but note that only its off-diagonal elements are relevant).

The inferred interaction weights \(w_{ij}\) are then renormalized as in Eq.\ref{renorm_eq}, which
in matrix notation reads
\begin{equation}
\tilde{\mathbf{W}^I}
=
\frac{\mathbf{W}^I}
{\mathbf{1}+\mathbf{A}+\mathbf{A}^{T}}.
\end{equation}

\begin{figure}
\centering
\includegraphics[width=0.475\linewidth]{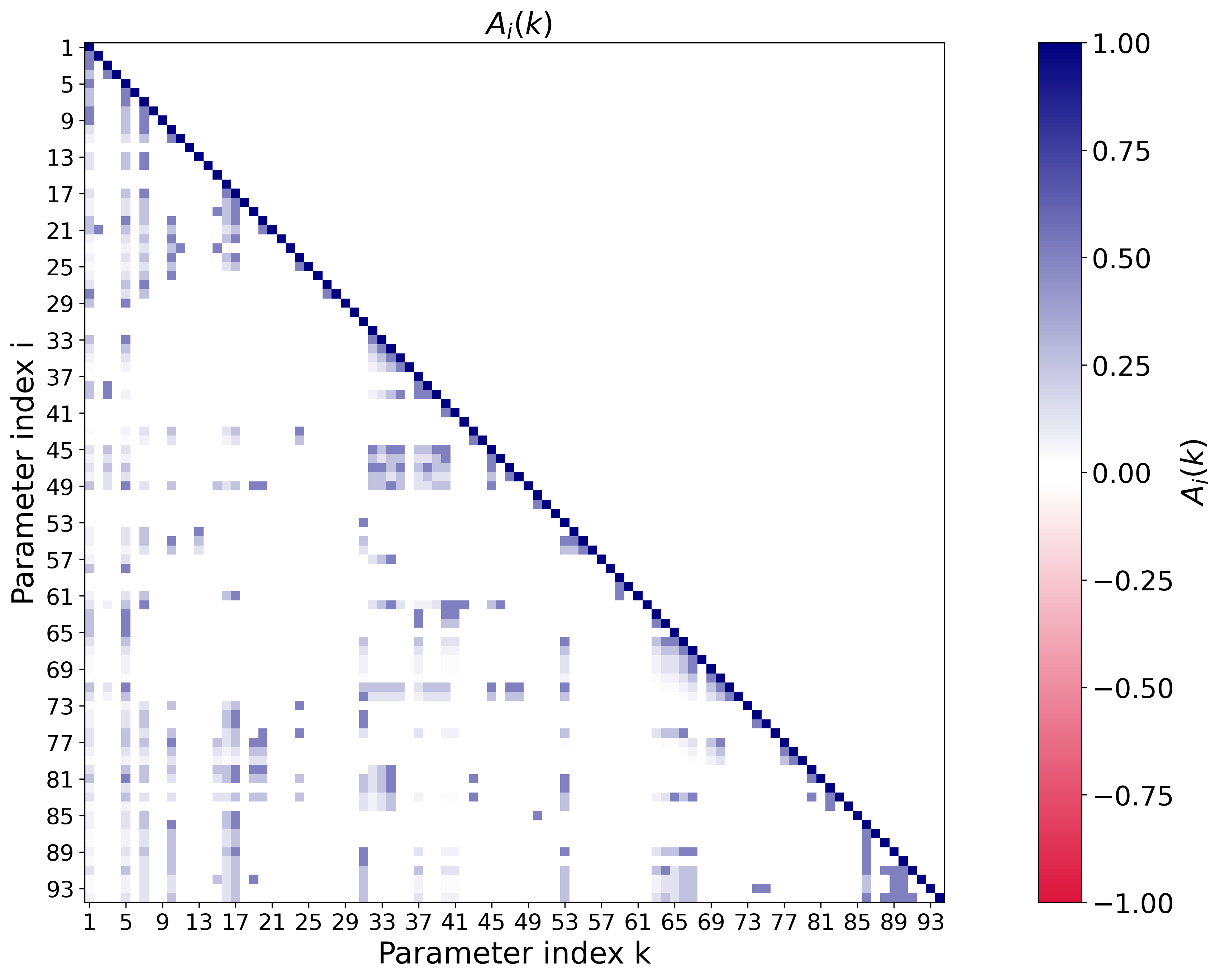}
\hfill
\includegraphics[width=0.475\linewidth]{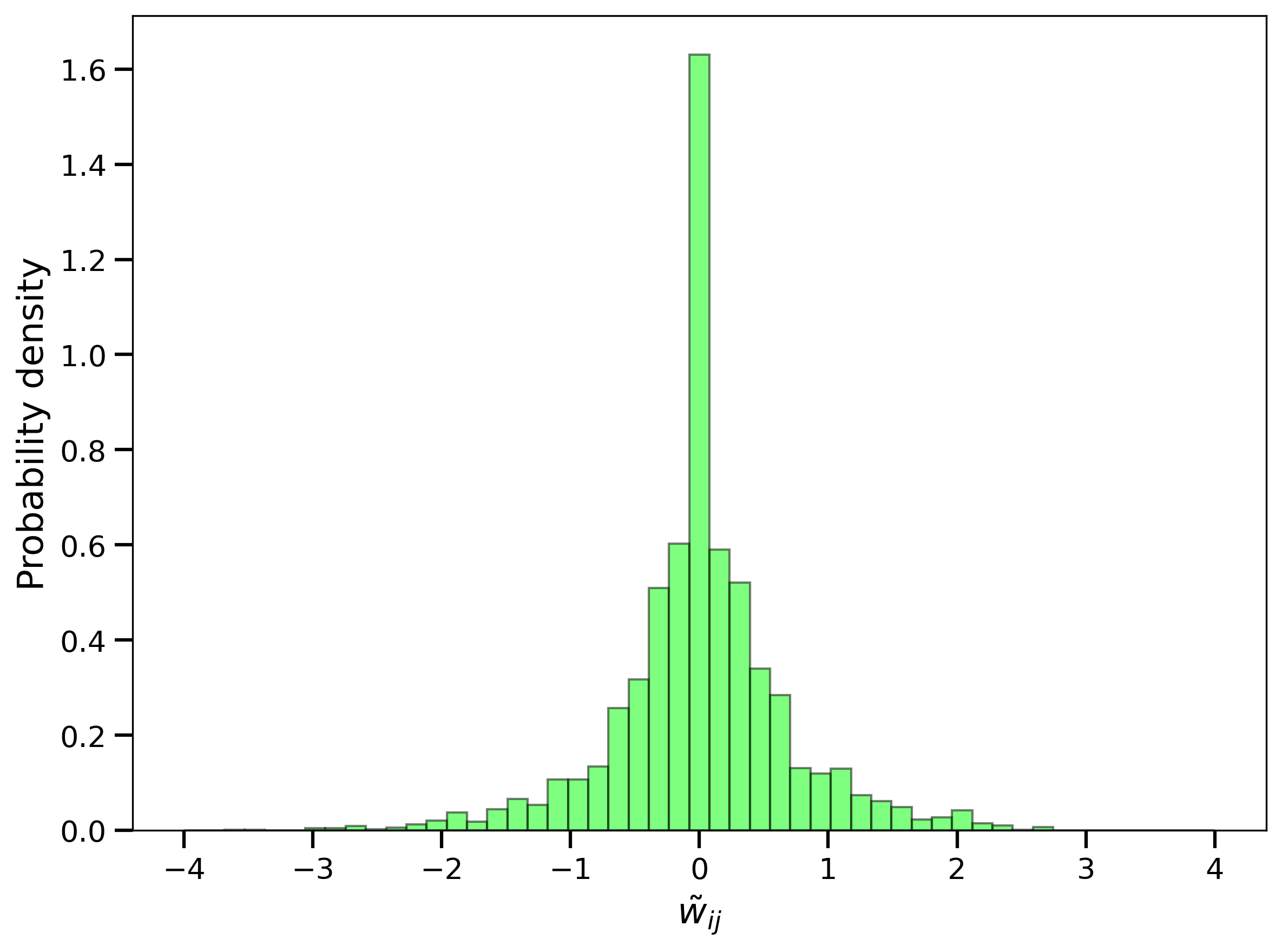}
\caption{Heatmap of renormalization terms $A^k_i$ used to renormalize the inferred weights, as described (left). Distributions of renormalized $\Tilde{w}^I_{ij}$ (right), to be compared to that in Fig.\ref{fig:alpha_comparison} (left).}
\label{fig:renorm}
\end{figure}

\bibliographystyle{unsrt} 
\bibliography{refs}
\end{document}